\documentclass[english]{article}
\usepackage[T1]{fontenc}
\usepackage[latin9]{inputenc}
\usepackage{geometry}
\usepackage{color}
\usepackage{array}
\usepackage{url}
\usepackage{multirow}
\usepackage{amsmath}
\usepackage{graphicx}
\usepackage{setspace}
\usepackage[all]{xy}
\makeatletter

\newcommand{\noun}[1]{\textsc{#1}}
\providecommand{\tabularnewline}{\\}

\newenvironment{lyxlist}[1]
	{\begin{list}{}
		{\settowidth{\labelwidth}{#1}
		 \setlength{\leftmargin}{\labelwidth}
		 \addtolength{\leftmargin}{\labelsep}
		 }}
	{\end{list}}

\@ifundefined{date}{}{\date{}}
\usepackage{amsthm}
\usepackage{multicol}
\usepackage{array,multirow,graphicx}
\usepackage{adjustbox}
\usepackage{flushend}
\PassOptionsToPackage{hyphens}{url}\usepackage{hyperref}
\usepackage[left]{lineno}
\usepackage[all]{xy}
\usepackage{authblk}

\hypersetup{%
    pdfborder = {0 0 0}
}

\providecommand{\tabularnewline}{\\}

\def\frontmatter@abstractheading{}

\renewcommand{\p@subsection}{}
\renewcommand{\p@subsubsection}{}

\usepackage{babel}
\usepackage{xcolor}
\usepackage{tikz}
\usepackage{adjustbox}

\makeatother

\usepackage{babel}
\begin{document}
\title{\textcolor{black}{True Contextuality Beats Direct Influences in Human
Decision Making}}
\author{\textcolor{black}{Irina Basieva}\textsuperscript{\textcolor{black}{1}}\textcolor{black}{,
Víctor H. Cervantes}\textsuperscript{\textcolor{black}{2}}\textcolor{black}{,\ \ Ehtibar
N. Dzhafarov}\textsuperscript{\textcolor{black}{3}}\textcolor{black}{,
Andrei Khrennikov}\textsuperscript{\textcolor{black}{4}}\textcolor{black}{\\}\textsuperscript{\textcolor{black}{1}}\textcolor{black}{City
University of London, UK (irina.basieva@gmail.com)\\ }\textsuperscript{\textcolor{black}{2}}\textcolor{black}{Purdue
University, USA (cervantv@purdue.edu)\\ }\textsuperscript{\textcolor{black}{3}}\textcolor{black}{Purdue
University, USA (ehtibar@purdue.edu)\\ }\textsuperscript{\textcolor{black}{4}}\textcolor{black}{Linnaeus
University, Sweden (andrei.khrennikov@lnu.se)}}
\maketitle
\begin{abstract}
\textcolor{black}{In quantum physics there are well-known situations
when measurements of the same property in different contexts (under
different conditions) have the same probability distribution, but
cannot be represented by one and the same random variable. Such systems
of random variables are called contextual. More generally, true contextuality
is observed when different contexts force measurements of the same
property (in psychology, responses to the same question) to be more
dissimilar random variables than warranted by the difference of their
distributions. The difference in distributions is itself a form of
context-dependence, but of another nature: it is attributable to direct
causal influences exerted by contexts upon the random variables. The
Contextuality-by-Default (CbD) theory allows one to separate true
contextuality from direct influences in the overall context-dependence.
The CbD analysis of numerous previous attempts to demonstrate contextuality
in human judgments shows that all context-dependence in them can be
accounted for by direct influences, with no true contextuality present.
However, contextual systems in human behavior can be found. In this
paper we present a series of crowdsourcing experiments that exhibit
true contextuality in simple decision making. The design of these
experiments is an elaboration of one introduced in the ``Snow Queen''
experiment (}\textcolor{black}{\emph{Decision}}\textcolor{black}{{}
5, 193-204, 2018), where contextuality was for the first time demonstrated
unequivocally.}

\textcolor{black}{KEYWORDS: concept combinations, context-dependence,
contextuality, direct influences. \\\\}
\end{abstract}

\section{\textcolor{black}{Introduction}}

\textcolor{black}{A response to a stimulus (say, a question) is generally
a }\textcolor{black}{\emph{random variable}}\textcolor{black}{{} that
can take on different values (say, Yes or No) with certain probabilities.
The }\textcolor{black}{\emph{identity}}\textcolor{black}{{} of a random
variable, in nontechnical terms, is what uniquely distinguishes this
random variable from other random variables.}\footnote{\textcolor{black}{In rigorous mathematical terms, a random variable
is defined as a (measurable) function mapping a domain probability
space into another (measurable) space. Its distribution is just one
property of this function, the probability measure it induces on the
codomain space.}}\textcolor{black}{{} The }\textcolor{black}{\emph{distribution}}\textcolor{black}{{}
of this random variable (probabilities with which it takes on different
values) is part of this identity, but clearly not the entire identity:
think of a handful of fair coins \textemdash{} a set of distinct random
variables with the same distribution. Other stimuli (e.g., other questions
posed together or prior to a given one) may }\textcolor{black}{\emph{directly
influence}}\textcolor{black}{{} the identity of the response to the
given stimulus by changing its distribution. In fact, this change
in the distribution, mathematically, is how the ``directness'' of
the influence is defined. True }\textcolor{black}{\emph{contextuality}}\textcolor{black}{{}
is such dependence of the identity of a response to a stimulus on
other stimuli that cannot be wholly explained by such direct influences.
We will elaborate this definition below. }

\textcolor{black}{Contextuality is at the very heart of quantum mechanics
(see, e.g., Liang, Spekkens, \& Wiseman, 2011), where it can be observed
by eliminating (or at least greatly reducing) all direct influences
by experimental design. (In quantum physics ``response to a stimulus''
has to be replaced with ``measurement of a property,'' but this
is in essence the same input-output relation.) This paper addresses
a question that ever since the 1990's interested researchers in physics,
computer science, and psychology, the question of whether true contextuality
can be observed outside quantum mechanics, with special interest (largely
for philosophical reasons we will not be discussing) in whether it
is present in human behavior. Many previous behavioral experiments
designed to answer this question (e.g., Aerts, 2014; Aerts, Gabora,
\& Sozzo, 2013; Asano, Hashimoto, Khrennikov, Ohya, \& Tanaka, 2014;
Bruza, Kitto, Nelson, \& McEvoy, 2009; Bruza, Kitto, Ramm, \& Sitbon,
2015; Bruza, Wang, \& Busemeyer, 2015; Cervantes \& Dzhafarov, 2017a,
2017b; Dzhafarov \& Kujala, 2014b; Dzhafarov, Kujala, Cervantes, Zhang,
\& Jones, 2016; Dzhafarov, Zhang, \& Kujala, 2015; Zhang \& Dzhafarov,
2017) have been shown to result in systems of random variables that
are noncontextual. This prompted Dzhafarov, Zhang, and Kujala (2015)
to consider the possibility that human behavior may never exhibit
true contextuality. It turns out, however, that contextual systems
in human behavior can be found. In this paper we describe a series
of experiments that, added to one previously conducted (Cervantes
\& Dzhafarov, 2018), demonstrate this unequivocally.}

\textcolor{black}{It should be emphasized at the outset that it would
be incorrect to think of contextuality as being ``surprising'' and
``strange'' while noncontextuality is ``trivial'' and ``expected.''
In the absence of constraints imposed by a general psychological theory,
comparable to quantum mechanics, we have no justification for such
judgements. One might argue in fact that it is most surprising that
so many experiments in psychology are described by noncontextual systems
of random variables. Nor would it be correct to assume that typical
psychological models, even very simple ones, can only predict noncontextual
systems: thus, in the concluding section of this paper we mention
a simple model that, on the contrary, predicts only contextual systems
(and has to be dismissed because of this). Contextuality analysis
is not a predictive model of behavior, and both contextual and noncontextual
systems are compatible with ``ordinary'' psychological models. In
that, as we point out in Section \ref{sec: Discussion}, psychology
is not different from quantum physics, where (non)contextuality of
a system is established based on the laws of quantum physics but is
not used to derive or revise them. What contextuality analysis elucidates
is the nature and structure of random variables \textemdash{} arguably,
the most basic and mandatory construct in the scientific analysis
of empirical systems, whether in psychology or elsewhere. In a well-defined
and mathematically rigorous sense, in a contextual system random variables
form true ``wholes'' that cannot be reduced to sets of distinct
random variables measuring or responding to specific elements of contexts
while being also cross-influenced by other elements of contexts. This
makes contextuality analysis inherently interesting, but we need much
greater knowledge of which behavioral systems are contextual and which
are not in order to determine what other properties of behavior these
characteristics are related to. We will return to the role and meaning
of contextuality after we introduce necessary definitions, theoretical
results, and empirical evidence. }

\subsection{\textcolor{black}{\label{subsec: Direct-influences-and}Direct influences
and true contextuality}}

\textcolor{black}{We introduce the basic notions related to contextuality
analysis using a simple example \textemdash{} responses to three Yes/No
questions asked two at a time. Most of the experiments reported below
are of this kind. Let, e.g., the three questions be}
\begin{lyxlist}{00.00.0000}
\item [{\textcolor{black}{$q_{1}$:}}] \textcolor{black}{Do you like chocolate?}
\item [{\textcolor{black}{$q_{2}$:}}] \textcolor{black}{Are you afraid
of pain?}
\item [{\textcolor{black}{$q_{3}$:}}] \textcolor{black}{Do you see your
dentist regularly?}
\end{lyxlist}
\textcolor{black}{Let a very large group of people be divided into
three subgroups: in the first subgroup each respondent is asked questions
$q_{1}$ and $q_{2}$; in the second subgroup each respondent is asked
questions $q_{2}$ and $q_{3}$; and in the third subgroup the questions
are $q_{3}$ and $q_{1}$. We call these pairwise arrangements of
questions }\textcolor{black}{\emph{contexts}}\textcolor{black}{, and
we denote them $c_{1},c_{2},c_{3}$, respectively. It does not matter
for the example whether the questions are asked in a fixed order,
randomized order, or (if in writing) simultaneously. A response to
question $q_{i}$ asked in context $c_{j}$ is a random variable that
we denote $R_{i}^{j}$: some of the people in the subgroup corresponding
to context $c_{j}$ will answer question $q_{i}$ with Yes, others
with No. Assuming the subgroups are so large that statistical issues
can be ignored, by counting the numbers of responses we can get a
good estimate of the probability distribution for our random variable:
\begin{equation}
R_{i}^{j}:\begin{array}{|c|c||c|}
\hline Yes & No & \textnormal{response}\\
\hline p_{i}^{j} & 1-p_{i}^{j} & \textnormal{probability}
\\\hline \end{array}.
\end{equation}
All in all we have six random variables in play, and they can be arranged
in the form of the following }\textcolor{black}{\emph{content-context}}\textcolor{black}{{}
matrix (Dzhafarov \& Kujala, 2016):
\begin{equation}
\begin{array}{|c|c|c||c|}
\hline R_{1}^{1} & R_{2}^{1} &  & c_{1}\\
\hline  & R_{2}^{2} & R_{3}^{2} & c_{2}\\
\hline R_{1}^{3} &  & R_{3}^{3} & c_{3}\\
\hline\hline q_{1} & q_{2} & q_{3} & \textnormal{system }\mathcal{R}_{3}
\\\hline \end{array}.\label{eq: matrix R3}
\end{equation}
}

\textcolor{black}{Now, the distributions of the responses to question
$q_{i}$ should be expected to differ depending on the context in
which it is asked. For instance, when $q_{1}$ (Do you like chocolate?)
is asked in combination with $q_{2}$ (Are you afraid of pain?), the
probability of $R_{1}^{1}=$``Yes, I like chocolate'' may be relatively
high, because chocolate is usually liked, and the mentioning of pain
in $q_{2}$ may make it sound especially comforting. However, when
the same question $q_{1}$ is asked in context $c_{3}$, in combination
with mentioning a dentist, the probability of $R_{1}^{3}=$``Yes,
I like chocolate'' may very well be lower. The same reasoning applies
to the two other questions: the responses to each of them will generally
be distributed differently depending on its context. This type of
influence exerted by a context on the responses to questions within
this context can be called }\textcolor{black}{\emph{direct influence}}\textcolor{black}{.
Indeed, the dependence of $R_{1}^{1}$ (responding to $q_{1}$) on
$q_{2}$ (another question in the same context) is essentially of
the same nature as the dependence of $R_{1}^{1}$ on $q_{1}$: a response
to $q_{1}$ is based on the information contained in $q_{1}$ }\textcolor{black}{\emph{and}}\textcolor{black}{{}
(even if to a lesser extent) on the information contained in $q_{2}$.
The other question in the same context can be viewed as part of the
question to which a response is given. }

\textcolor{black}{Is all context-dependence of this direct influence
variety? As it turns out, the answer is negative. Imagine, e.g., that
all direct influences are eliminated by some procedural trick, and
each question in each context is answered Yes with probability $1/2$.
This means, in particular, that $R_{1}^{1}$ and $R_{1}^{3}$ have
one and the same distribution,
\begin{equation}
R_{1}^{1}:\begin{array}{|c|c|}
\hline Yes & No\\
\hline 1/2 & 1/2
\\\hline \end{array},\:R_{1}^{3}:\begin{array}{|c|c|}
\hline Yes & No\\
\hline 1/2 & 1/2
\\\hline \end{array},
\end{equation}
and if one does not take into account their relations to $R_{2}^{1}$
(in context $c_{1}$) and to $R_{3}^{3}$ (in context $c_{3}$), one
could consider $R_{1}^{1}$ and $R_{1}^{3}$ as if they were always
equal to each other \textemdash{} essentially one and the same random
variable.}\footnote{\textcolor{black}{The ``as if'' here serves to circumvent the technicalities
associated with the fact that, strictly speaking, we are dealing here
not with $R_{1}^{1}$ and $R_{1}^{3}$ themselves but with their probabilistic
copies (}\textcolor{black}{\emph{couplings}}\textcolor{black}{) that
are jointly distributed. See Dzhafarov and Kujala (2014a, 2017b) for
details.}}\textcolor{black}{{} And similarly for $R_{2}^{1}$ and $R_{2}^{2}$,
and for $R_{3}^{2}$ and $R_{3}^{3}$. If one looks at each column
of matrix (\ref{eq: matrix R3}) separately, ignoring the row-wise
joint distributions, then one can write 
\begin{equation}
\begin{array}{c}
R_{1}^{1}=R_{1}^{3}\\
R_{2}^{1}=R_{2}^{2}\\
R_{3}^{2}=R_{3}^{3}
\end{array}.\label{eq: identity coupling}
\end{equation}
Consider, however, the possibility that no respondent ever gives the
same answer to both questions posed to her. Thus, if she answers Yes
to $q_{1}$ in context $c_{1}$ (which can happen with probability
$1/2$), she always answers No to $q_{2}$, and vice versa. Denoting
Yes and No by $+1$ and $-1$, respectively, we have a chain of equalities
\begin{equation}
\begin{array}{c}
R_{1}^{1}=-R_{2}^{1}\\
R_{2}^{2}=-R_{3}^{2}\\
R_{1}^{3}=-R_{3}^{3}
\end{array},\label{eq: +--}
\end{equation}
and it is clear that (\ref{eq: identity coupling}) and (\ref{eq: +--})
cannot be satisfied together: combining them would lead to a numerical
contradiction. We should conclude therefore that when the joint distributions
within contexts are taken into account, $R_{1}^{1}$ and $R_{1}^{3}$,
or $R_{2}^{1}$ and $R_{2}^{2}$, or $R_{3}^{2}$ and $R_{3}^{3}$
cannot be considered always equal to each other. In at least one of
these pairs, the two random variables should be more different than
it is warranted by their individual distributions (which are, in this
example, identical). This is a situation in which we can say that
the system exhibits }\textcolor{black}{\emph{true contextuality}}\textcolor{black}{,
the kind of context-dependence that is not reducible to direct influences
(in this example, absent). }

\textcolor{black}{Empirical data, especially outside quantum physics,
almost always involve some direct influences, but the logic of finding
out whether they also involve true contextuality remains the same.
Continuing to use matrix (\ref{eq: matrix R3}) as a demonstration
tool, we first look at the columns of the matrix one by one, ignoring
the contexts. For each pair of random variables in a column (responses
to the same question), we find out how close to each other they could
be made if they were jointly distributed. In other words, we find
the maximal probabilities with which each of the equalities in (\ref{eq: identity coupling})
can be satisfied. Then we investigate whether all the variables in
our system can be made jointly distributed while preserving these
maximal probabilities. If the answer is negative, we conclude that
the contexts force the random variables sharing a column to be more
dissimilar than warranted by direct influences (differences in their
individual distributions). We then call such a system }\textcolor{black}{\emph{contextual}}\textcolor{black}{.
Otherwise it is }\textcolor{black}{\emph{noncontextual}}\textcolor{black}{.
This is the gist of the approach to contextuality called Contextuality-by-Default
(CbD), and we illustrate it in the next section by a detailed numerical
example. }

\textcolor{black}{CbD forms the theoretical basis for the design and
analysis of our experiments. For completeness, however, another approach
to the notion of contextuality should be mentioned, one treating context-dependent
probabilities as a generalization of conditional probabilities defined
through Bayes's formula (Khrennikov, 2009). With some additional assumptions
these contextual probabilities can be represented by quantum-theoretical
formalisms \textemdash{} state vectors in complex Hilbert space and
Hermitian operators or their generalizations. Applications of such
approach to cognitive psychology can be found in Khrennikov (2010)
and Busemeyer and Bruza (2012), among other monographs and papers.
CbD, by contrast, is squarely within classical probability theory.
Although contextuality in CbD can be called ``quantum-like'' due
to the origins of the concept in quantum physics, CbD uses no quantum
formalisms.}

\subsection{\textcolor{black}{\label{subsec: A-numerical-example}A numerical
example and interpretation}}

\textcolor{black}{The following numerical example illustrates how
CbD works. Let there be just two dichotomous questions, $q_{1}$ and
$q_{2}$, answered in two contexts, $c_{1}$ and $c_{2}$ (e.g., in
two different orders, as in Wang \& Busemeyer, 2013). The content-context
matrix here is
\begin{equation}
\begin{array}{|c|c||c|}
\hline R_{1}^{1} & R_{2}^{1} & c_{1}\\
\hline R_{1}^{2} & R_{2}^{2} & c_{2}\\
\hline\hline q_{1} & q_{2} & \textnormal{system }\mathcal{R}_{2}
\\\hline \end{array}.\label{eq: matrix R2}
\end{equation}
Assume that the joint distributions along the rows of the matrix are
as shown:}

\textcolor{black}{
\begin{equation}
\begin{array}{l|c|c|c}
c_{1} & R_{2}^{1}=Yes & R_{2}^{1}=No\\
\hline R_{1}^{1}=Yes & 1/2 & 0 & 1/2\\
\hline R_{1}^{1}=No & 0 & 1/2 & 1/2\\
\hline  & 1/2 & 1/2
\end{array}\quad\begin{array}{l|c|c|c}
c_{2} & R_{1}^{2}=Yes & R_{1}^{2}=No\\
\hline R_{2}^{2}=Yes & a & 1/2-a & 1/2\\
\hline R_{2}^{2}=No & 3/4-a & a-1/4 & 1/2\\
\hline  & 3/4 & 1/4
\end{array},\label{eq: two matrices}
\end{equation}
where $a$ is some value between $1/4$ and $1/2$. Knowing these
distributions means that, for any filling of the matrix (\ref{eq: matrix R2})
with values of the random variables $R_{1}^{1},R_{2}^{1},R_{1}^{2},R_{2}^{2}$
(Yes or No, for a total of 16 combinations), we know the row-wise
probabilities: e.g., 
\begin{equation}
\begin{array}{|c|c||c|}
\hline R_{1}^{1}=Yes & R_{2}^{1}=Yes & p_{1}\left(Yes,Yes\right)=1/2\\
\hline R_{1}^{2}=Yes & R_{2}^{2}=No & p_{2}\left(Yes,No\right)=3/4-a
\\\hline \end{array}.\label{eq: rowwise}
\end{equation}
We see from (\ref{eq: two matrices}) that $R_{2}^{1}$ and $R_{2}^{2}$
(the responses to question $q_{2}$) are distributed identically.
Because of this, if they were jointly distributed (see footnote 1),
the maximal probability with which they could be equal to each other
would be 1: 
\begin{equation}
\begin{array}{l|c|c|c}
q_{2} & R_{2}^{2}=Yes & R_{2}^{2}=No\\
\hline R_{2}^{1}=Yes & 1/2 & 0 & 1/2\\
\hline R_{2}^{1}=No & 0 & 1/2 & 1/2\\
\hline  & 1/2 & 1/2
\end{array}.\label{eq: q2}
\end{equation}
The responses to question $q_{1}$, however, are distributed differently,
and in the imaginary matrix of their joint distribution,
\begin{equation}
\begin{array}{l|c|c|c}
q_{1} & R_{1}^{2}=Yes & R_{1}^{2}=No\\
\hline R_{1}^{1}=Yes & 1/2 & 0 & 1/2\\
\hline R_{1}^{1}=No & 1/4 & 1/4 & 1/2\\
\hline  & 3/4 & 1/4
\end{array},\label{eq: q1}
\end{equation}
the maximal possible probability of $R_{1}^{1}=R_{1}^{2}=Yes$ is
$1/2$, and the maximal possible value of $R_{1}^{1}=R_{1}^{2}=No$
is $1/4$. Therefore, if they were jointly distributed, the maximal
probability with which $R_{1}^{1}=R_{1}^{2}$ would be 3/4. Now, with
these imaginary distributions, for any filling of the matrix (\ref{eq: matrix R2})
with Yes-No values of the random variables $R_{1}^{1},R_{2}^{1},R_{1}^{2},R_{2}^{2}$,
we also have the column-wise probabilities: e.g., 
\begin{equation}
\begin{array}{|c|c|}
\hline R_{1}^{1}=Yes & R_{2}^{1}=Yes\\
\hline R_{1}^{2}=Yes & R_{2}^{2}=No\\
\hline p'_{1}\left(Yes,Yes\right)=1/2 & p'_{2}\left(Yes,No\right)=0
\\\hline \end{array}.\label{eq: columnwise}
\end{equation}
The problem we have to solve now is: are these column-wise probabilities
compatible with the row-wise probabilities in (\ref{eq: rowwise})?
The compatibility means that, to any of the 16 filling of the matrix
(\ref{eq: matrix R2}) with values of the random variables $R_{1}^{1},R_{2}^{1},R_{1}^{2},R_{2}^{2}$,
we can assign a probability, e.g., $p\left(R_{1}^{1}=Yes,R_{2}^{1}=Yes,R_{1}^{2}=Yes,R_{2}^{2}=No\right)$,
such that the row-wise sums of these probabilities agree with (\ref{eq: rowwise})
and the column-wise sums agree with (\ref{eq: columnwise}). This
is a classical linear programming problem: for any given value of
$a$ it is guaranteed that either such an assignment of probabilities
will be found (so that the system is noncontextual) or the determination
will be made that such an assignment does not exist (the system is
contextual). In our case, however, one need not resort to linear programing
to see that no such assignment of probabilities is possible for any
value of $a$ other than $1/2$. Indeed, we see from the $c_{1}$-distribution
in (\ref{eq: two matrices}) and from (\ref{eq: q2}) that, with probability
1,}

\textcolor{black}{
\begin{equation}
R_{1}^{1}=R_{2}^{1}=R_{2}^{2}.\label{eq: triple}
\end{equation}
So, $R_{1}^{1}$ and $R_{2}^{2}$ are essentially the same random
variable, say $X$. But, from (\ref{eq: q1}), this $X$ equals $R_{1}^{2}$
with probability $3/4$, whereas from the $c_{2}$-distribution in
(\ref{eq: two matrices}), this $X$ equals $R_{1}^{2}$ with probability
$2a-1/4$, which is not $3/4$ if $a\not=1/2$. The conclusion is
that the joint distributions along the two rows of the content-context
matrix (\ref{eq: matrix R2}) prevent the responses to the same questions
in the two columns of the matrix to be as close to each other as they
can be if the two columns are viewed separately. The system therefore
is contextual for any $a\not=1/2$. }

\textcolor{black}{Why is this interesting? In psychological terms,
the interpretation of the question order effect seems straightforward:
the first question reminds something or draws one's attention to something
that is relevant to the second question. What is shown by the contextuality
analysis of our hypothetical question-order system is that this interpretation
is only sufficient for $a=1/2$, being incomplete in all other cases.
The responses to two questions posed in a particular order form a
``whole'' that cannot be reduced to an action of the first question
upon the second response: the identity of the two random variables
changes beyond the effect of this action on their distributions. We
will return to this issue in the concluding section of the paper.}

\textcolor{black}{The reader should not forget that we are discussing
a numerical example rather than experimental data. The large body
of experimental data on the question-order effect collected by Wang
and Busemeyer (2013) has been subjected to contextual analysis in
Dzhafarov, Zhang, and Kujala (2015), the result being that the responses
to any of the many pairs of questions studied exhibit no contextuality.
In fact, almost all question pairs are in a good agreement with the
``QQ law'' discovered by Wang and Busemeyer (2013), 
\begin{equation}
\Pr\left[R_{1}^{1}=R_{2}^{1}\right]=\Pr\left[R_{1}^{2}=R_{2}^{2}\right],
\end{equation}
and, as shown in Dzhafarov, Zhang, and Kujala (2015), this law implies
no contextuality: this system of random variables is entirely describable
in terms of each response being dependent on ``its own'' question,
plus the second respond being also influenced by the first question.
The idea of a ``whole'' being irreducible to interacting parts is
not therefore an automatically applicable formula. To see if it is
applicable at all, in psychology, one should look for empirical evidence
elsewhere. Such evidence is presented below.}

\subsection{\textcolor{black}{\label{subsec: Contetuality-by-Default}Contextuality-by-Default}}

\textcolor{black}{CbD was developed (Dzhafarov, Cervantes, Kujala,
2017; Dzhafarov \& Kujala 2014a, 2016, 2017a, 2017b; Kujala, Dzhafarov,
\& Larsson, 2015) as a generalization of the quantum-mechanical notion
of contextuality (Abramsky \& Brandenburger, 2011; Fine, 1982; Kochen
\& Specker, 1967; Kurzynski, Ramanathan, \& Kaszlikowski, 2012). The
latter only applies to }\textcolor{black}{\emph{consistently connected}}\textcolor{black}{{}
systems, those in which direct influences are absent, i.e., responses
to the same stimulus (or measurements of the same property) in different
contexts are distributed identically. In physics this requirement
is known by such names as ``no-signaling,'' ``no-disturbance,''
etc.; in psychology it is known as }\textcolor{black}{\emph{marginal
selectivity}}\textcolor{black}{{} (Dzhafarov, 2003; Townsend \& Schweickert,
1989). This requirement is never satisfied in behavioral experiments
(Dzhafarov \& Kujala, 2014b; Dzhafarov, Kujala, Cervantes, Zhang,
\& Jones, 2016; Dzhafarov, Zhang, \& Kujala, 2015), and it is often
violated in quantum physical experiments too (Adenier \& Khrennikov,
2017; Kujala, Dzhafarov, \& Larsson, 2015). The main difficulty faced
by many previous attempts to reveal contextuality in human behavior
was that they could not apply mathematical tests predicated on the
assumption of consistent connectedness to systems in which this requirement
does not hold. As mentioned in the introduction, a CbD-based analysis
of these experiments (Dzhafarov \& Kujala, 2014b; Dzhafarov, Kujala,
Cervantes, Zhang, \& Jones, 2016; Dzhafarov, Zhang, \& Kujala, 2015)
showed that all context-dependence in them was attributable to direct
influences. The first unequivocal evidence of the existence of contextual
systems in human behavior was provided by Cervantes and Dzhafarov's
(2018) ``Snow Queen'' experiment. }

\textcolor{black}{The idea underlying the design of the ``Snow Queen''
experiment (and all the experiments reported below) is suggested by
the criterion (necessary and sufficient condition) of contextuality
when CbD is applied to cyclic systems with dichotomous random variables
(Dzhafarov, Kujala, \& Larsson, 2015; Kujala \& Dzhafarov, 2016; Kujala,
Dzhafarov, \& Larsson, 2015). In such a system $n$ questions and
$n$ contexts can be arranged as
\begin{equation}
\xymatrix@C=1cm{q_{1}\ar@{-}[r]^{c_{1}} & q_{2}\ar@{-}[r]^{c_{2}} & \cdots\ar@{-}[r]^{c_{n-2}} & q_{n-1}\ar@{-}[r]^{c_{n-1}} & q_{n}\ar@{-}@/^{1pc}/[llll]^{c_{n}}}
\label{eq: cycle n}
\end{equation}
The number $n$ is referred to as the rank of the system. The question-order
system (\ref{eq: matrix R2}) considered in Section \ref{subsec: A-numerical-example}
is the smallest possible cyclic system, of rank 2,
\begin{equation}
\xymatrix@C=1cm{q_{1}\ar@{-}[r]^{c_{1}} & q_{2}\ar@{-}@/^{1pc}/[l]^{c_{2}}.}
\end{equation}
The system (\ref{eq: matrix R3}) in Section \ref{subsec: Direct-influences-and}
is a cyclic system of rank 3,
\begin{equation}
\xymatrix@C=1cm{q_{1}\ar@{-}[r]^{c_{1}} & q_{2}\ar@{-}[r]^{c_{2}} & q_{3}\ar@{-}@/^{1pc}/[ll]^{c_{3}},}
\end{equation}
and it is used in four of the six experiments reported below. The
remaining two are analyzed as cyclic systems of rank 4, 
\begin{equation}
\xymatrix@C=1cm{q_{1}\ar@{-}[r]^{c_{1}} & q_{2}\ar@{-}[r]^{c_{2}} & q_{3}\ar@{-}[r]^{c_{n-1}} & q_{4}\ar@{-}@/^{1pc}/[lll]^{c_{n}}}
,
\end{equation}
with the content-context matrix 
\begin{equation}
\begin{array}{|c|c|c|c||c|}
\hline R_{1}^{1} & R_{2}^{1} & {\color{blue}} & {\color{blue}} & c_{1}\\
\hline {\color{blue}} & R_{2}^{2} & R_{3}^{2} & {\color{blue}} & c_{2}\\
\hline {\color{blue}} & {\color{blue}} & R_{3}^{3} & R_{4}^{3} & c_{3}\\
\hline R_{1}^{4} & {\color{blue}} & {\color{blue}} & R_{4}^{4} & c_{4}\\
\hline\hline q_{1} & q_{2} & q_{3} & q_{4} & \textnormal{system }\mathcal{R}_{4}
\\\hline \end{array}.\label{eq: matrix R4}
\end{equation}
}

\textcolor{black}{To formulate the criterion of contextuality in cyclic
systems, we encode the values of our random variables by $+1$ and
$-1$. Then the products of the random variables in the same context,
such as $R_{1}^{1}R_{2}^{1}$, are well-defined, and so are the expected
values $\mathsf{E}\left[R_{1}^{1}R_{2}^{1}\right]$, $\mathsf{E}\left[R_{2}^{2}R_{3}^{2}\right]$,
etc. For instance, if the joint distribution of $R_{1}^{1}$ and $R_{2}^{1}$
(responses to questions $q_{1}$ and $q_{2}$ in context $c_{1}$)
is
\begin{equation}
\begin{array}{l|c|c|c}
c_{1} & R_{2}^{1}=+1 & R_{2}^{1}=-1\\
\hline R_{1}^{1}=+1 & a & b & a+b\\
\hline R_{1}^{1}=-1 & c & d & c+d\\
\hline  & a+c & b+d
\end{array},
\end{equation}
then $R_{1}^{1}R_{2}^{1}$ has the distribution
\begin{equation}
\begin{array}{|c|c|}
R_{1}^{1}R_{2}^{1}=+1 & R_{1}^{1}R_{2}^{1}=-1\\
\hline a+d & b+c
\\\hline \end{array},
\end{equation}
and the distribution of $R_{1}^{1}$ and $R_{2}^{1}$ is described
by the expected values 
\begin{equation}
\begin{array}{l}
\mathsf{E}\left[R_{1}^{1}\right]=\left(a+b\right)-\left(c+d\right),\\
\mathsf{E}\left[R_{2}^{1}\right]=\left(a+c\right)-\left(b+d\right),\\
\mathsf{E}\left[R_{1}^{1}R_{2}^{1}\right]=\left(a+d\right)-\left(b+c\right).
\end{array}
\end{equation}
We will also need a special function, $s_{odd}$: given some real
numbers $x_{1},\ldots,x_{n}$, 
\begin{equation}
s_{odd}\left(x_{1},\ldots,x_{n}\right)=\max\left(\pm x_{1}\pm\ldots\pm x_{n}\right),
\end{equation}
where each $\pm$ is to be replaced with $+$ or $-$, and the maximum
is taken over all choices that contain an odd number of minus signs.
Thus, 
\begin{equation}
\begin{array}{c}
s_{odd}\left(x,y\right)=\max\left(-x+y,x-y\right),\\
s_{odd}\left(x,y,z\right)=\max\left(-x+y+z,x-y+z,x+y-z,-x-y-z\right)\\
\textnormal{etc.}
\end{array},
\end{equation}
}

\textcolor{black}{The theorem proved by Kujala and Dzhafarov (2016)
says that a cyclic system of rank $n$ is contextual (exhibits true
contextuality) if and only if
\begin{equation}
D=s_{odd}\left(\mathsf{E}\left[R_{1}^{1}R_{2}^{1}\right],\mathsf{E}\left[R_{2}^{2}R_{3}^{2}\right],\ldots,\mathsf{E}\left[R_{n}^{n}R_{1}^{n}\right]\right)-\left(n-2\right)-\Delta>0,\label{eq: criterion}
\end{equation}
where
\begin{equation}
\Delta=\left|\mathsf{E}\left[R_{1}^{1}\right]-\mathsf{E}\left[R_{1}^{n}\right]\right|+\left|\mathsf{E}\left[R_{2}^{1}\right]-\mathsf{E}\left[R_{2}^{2}\right]\right|+\ldots+\left|\mathsf{E}\left[R_{n}^{n-1}\right]-\mathsf{E}\left[R_{n}^{n}\right]\right|.\label{eq: delta}
\end{equation}
The value of $\Delta$ is a measure of direct influences, or of inconsistent
connectedness. It shows how much, overall, the distributions of responses
to one and the same question differ in different contexts. If $\Delta=0$,
the system is consistently connected: the response to a given question
is not influenced by the other questions with which it co-occurs in
the same context.}\footnote{\textcolor{black}{The special case of (\ref{eq: criterion}) for $\Delta=0$
was proved, by very different mathematical means, in Araújo, Quintino,
Budroni, Cunha, \& Cabello (2013).}}\textcolor{black}{{} One can loosely interpret $s_{odd}$ as a measure
of the ``potential true contextuality'': it shows how much, overall,
the identities of the random variables responding to the same question
differ in different contexts. The contextuality test for a cyclic
system therefore can be viewed as a test of whether these differences
exceed those due to direct influences alone. The failure of the previous
attempts to find contextuality in behavioral data may be described
by saying that the empirical situations chosen for investigation had
too strong direct influences for the amount of potential true contextuality
they contained.}

\textcolor{black}{The idea of the ``Snow Queen'' experiment was
to make the value of $s_{odd}$ as large as possible, increasing its
chances of ``beating'' $\Delta$, a quantity that cannot be controlled
by experimental design.}\footnote{\textcolor{black}{In physics the situation is different: one can eliminate
or greatly reduce direct influences by, e.g., separating two entangled
particles by a space-time interval that prevents transmission of a
signal between them.}}\textcolor{black}{{} The formal structure of the experiment was a cyclic
system of rank 4, with $q_{1}$ and $q_{3}$ being two choices of
characters from a story (Snow Queen, by H.C. Andersen), and $q_{2}$
and $q_{4}$ being two choices of attributes of these characters.
\begin{equation}
\begin{array}{|c|c|c|c||c|}
\hline R_{1}^{1} & R_{2}^{1} &  &  & c_{1}\\
\hline  & R_{2}^{2} & R_{3}^{2} &  & c_{2}\\
\hline  &  & R_{3}^{3} & R_{4}^{3} & c_{3}\\
\hline R_{1}^{4} &  &  & R_{4}^{4} & c_{4}\\
\hline\hline q_{1}:\begin{array}{c}
\textnormal{Gerda}\\
\textnormal{Troll}
\end{array} & q_{2}:\begin{array}{c}
\textnormal{beautiful}\\
\textnormal{unattractive}
\end{array} & q_{3}:\begin{array}{c}
\textnormal{Snow Queen}\\
\textnormal{old Finn woman}
\end{array} & q_{4}:\begin{array}{c}
\textnormal{kind}\\
\textnormal{evil}
\end{array} & \textnormal{system }\mathcal{SQ}_{4}
\\\hline \end{array}.
\end{equation}
For instance, in context $c_{3}$, a respondent could choose either
Snow Queen or old Finn woman, and also choose either ``kind'' or
``evil.'' The instruction said the choices had to match the story
line. The respondents knew, e.g., that Snow Queen is beautiful and
evil, and that the old Finn woman is unattractive and kind.}\footnote{\textcolor{black}{This instruction is an analogue of the quantum-mechanical
}\textcolor{black}{\emph{preparation}}\textcolor{black}{, an empirical
procedure preceding an experiment with the aim of creating a specific
pattern of high correlations between measurements.}}\textcolor{black}{{} It is easy to show that if all respondents followed
the instruction correctly, $s_{odd}$ in this experiment had to have
the maximal possible value of 4. The amount of direct influences measured
by $\Delta$ was considerable, but the left-hand side expression in
(\ref{eq: criterion}) was well above zero, with very high statistical
reliability (evaluated by 99.99\% bootstrap confidence intervals). }

\textcolor{black}{One possible criticism of the ``Snow-Queen'' experiment
can be that the paired choices were too ``asymmetric'': choice of
a character, such as Gerda, and choice of a characteristic, such as
``beautiful,'' seem too different in nature. In the experiments
reported below the paired choices were ``on a par.'' Otherwise,
the experiments followed the same logic, ensuring the highest possible
value for $s_{odd}$. This value equals $n$, the rank of the cyclic
system. In quantum physics, the systems with this property (if, additionally,
they are consistently connected, i.e., $\Delta=0$), are called PR-boxes,
after Popescu and Rohrlich (1994). In our experiments $n$ was 3 or
4. }

\section{\textcolor{black}{Method}}

\subsection*{\textcolor{black}{Participants}}

\textcolor{black}{We recruited 6192 participants on CrowdFlower (2018)
between February 7 and 12, 2018. They agreed to participate in this
study by accepting a standard consent from. The consent form and the
interactive experimental procedure were provided via a Qualtrics survey
hosted by City University London. The study was approved by City University
London Research Ethics Committee, PSYETH (S/L) 17/18 09. (The number
of participants was chosen so that we could construct reliable 99.99\%
bootstrap confidence intervals for each context in each experiment,
as described below.)}

\subsection*{\textcolor{black}{Materials and procedure}}

\textcolor{black}{Each respondent participated in all six experiments,
in a random order. For each of the experiments, each participant was
randomly and independently assigned to one of the conditions (contexts).
In each context, a participant was introduced to a pair of choices
to be made by a fictional Alice; each choice was between two alternatives.
There were three contexts in Experiments 1-4, and four contexts in
Experiments 5 and 6. Figure \ref{fig:appearance} shows the way the
instruction and choices were presented to respondents in one context
of Experiment 1.}
\begin{center}
\textcolor{black}{}
\begin{figure}
\begin{centering}
\textcolor{black}{\includegraphics[scale=0.6]{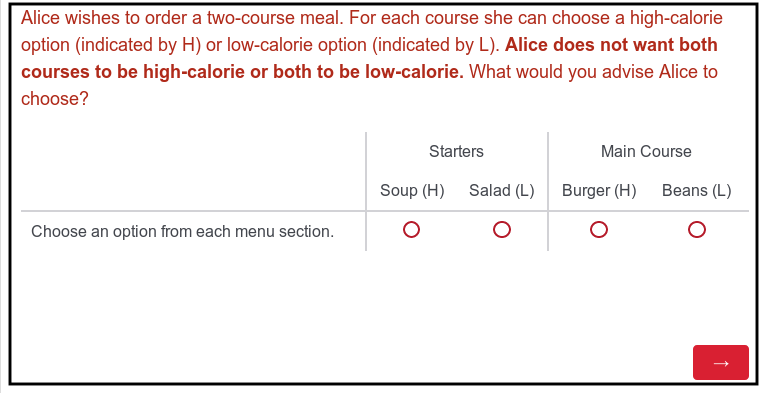}}
\par\end{centering}
\centering{}\textcolor{black}{\caption{The appearance of the computer screen to the participant if assigned
to context $c_{1}$ in experiment 1. The participant was required
to choose an option for each question, in this case each menu section;
the next experiment or the end of the survey would be reached by clicking
the `Next' arrow. If the participant had made both choices in accordance
with the instructions, in this case having chosen Soup (H) with Beans
(L) or Salad (L) with Burger (H), clicking the 'Next' arrow allowed
the survey to continue; otherwise the participants were prompted to
revise or complete their responses.\label{fig:appearance}}
}
\end{figure}
\par\end{center}

\subsubsection*{\textcolor{black}{Experiments 1 to 4}}

\textcolor{black}{In experiments 1-4, in each context, the character
Alice was faced with two choices out of a set of three dichotomous
choices. The participant was asked to select a pair of responses that
respected Alice's preferences as stated in the instructions (see Fig.
\ref{fig:appearance}). The system would not allow the respondent
to make only one choice or two choices contradicting the instructions.
The following depicts the situations presented, while table \ref{tab:Dichotomous-choices-1-4}
summarizes the sets of dichotomous choices.}
\begin{description}
\item [{\textcolor{black}{Experiment\ ``Meals.''}}] \textcolor{black}{Alice
wishes to order a two-course meal. For each course she can choose
a high-calorie option (indicated by H) or a low-calorie option (indicated
by L). Alice does }\textcolor{black}{\noun{not}}\textcolor{black}{{}
want both courses to be high-calorie nor does she want both of them
to be low-calorie.}
\item [{\textcolor{black}{Experiment\ ``Clothes.''}}] \textcolor{black}{Alice
is dressing for work, and chooses two pieces of clothing. She does
}\textcolor{black}{\noun{not}}\textcolor{black}{{} want both of them
to be plain, nor does she want both of them to be fancy.}
\item [{\textcolor{black}{Experiment\ ``Presents.''}}] \textcolor{black}{Alice
wishes to buy two presents for her nephew's birthday. She can choose
either a more expensive option (indicated by E) or a cheaper option
(indicated by C). Alice does }\textcolor{black}{\noun{not}}\textcolor{black}{{}
want both presents to be expensive or both presents to be cheap.}
\item [{\textcolor{black}{Experiment\ ``Exercises.''}}] \textcolor{black}{Alice
is doing two physical exercises. Alice does }\textcolor{black}{\noun{not}}\textcolor{black}{{}
want both exercises to be hard or both to be easy.}
\end{description}
\begin{center}
\textcolor{black}{}
\begin{table}
\begin{centering}
\textcolor{black}{\caption{Dichotomous choices in experiments 1 to 4.\textcolor{blue}{{} }\textcolor{black}{Each
respondent was asked to make two choices ($q_{1}\&q_{2}$ or $q_{2}\&q_{3}$
or $q_{3}\&q_{1}$), randomly and independently assigned to this respondent
in each experiment.}\label{tab:Dichotomous-choices-1-4}}
}%
\begin{tabular}{|l|c|c|c|}
\hline 
 & \textcolor{black}{$q_{1}$} & \textcolor{black}{$q_{2}$} & \textcolor{black}{$q_{3}$}\tabularnewline
\hline 
\hline 
\multirow{2}{*}{\textcolor{black}{1. Meals}} & \textcolor{black}{Starters: } & \textcolor{black}{Main course: } & \textcolor{black}{Dessert: }\tabularnewline
 & \textcolor{black}{Soup (H){*} or Salad (L)} & \textcolor{black}{Burger (H){*} or Beans (L)} & \textcolor{black}{Cake (H){*} or Coffee (L)}\tabularnewline
\hline 
\multirow{2}{*}{\textcolor{black}{2. Clothes}} & \textcolor{black}{Skirt:} & \textcolor{black}{Blouse:} & \textcolor{black}{Jacket:}\tabularnewline
 & \textcolor{black}{Plain{*} or Fancy} & \textcolor{black}{Plain{*} or Fancy} & \textcolor{black}{Plain{*} or Fancy}\tabularnewline
\hline 
\multirow{2}{*}{\textcolor{black}{3. Presents}} & \textcolor{black}{Book:} & \textcolor{black}{Soft toy (bear):} & \textcolor{black}{Construction set:}\tabularnewline
 & \textcolor{black}{Big expensive book (E){*} or Smaller book(C)} & \textcolor{black}{(E){*} or (C)} & \textcolor{black}{(E){*} or (C)}\tabularnewline
\hline 
\multirow{2}{*}{\textcolor{black}{4. Exercises}} & \textcolor{black}{Arms:} & \textcolor{black}{Back:} & \textcolor{black}{Legs:}\tabularnewline
 & \textcolor{black}{Hard{*} or Easy} & \textcolor{black}{Hard{*} or Easy} & \textcolor{black}{Hard{*} or Easy}\tabularnewline
\hline 
\end{tabular}
\par\end{centering}
\raggedright{}\textcolor{black}{{*} Denotes the response encoded with
$+1$}
\end{table}
\par\end{center}

\begin{center}
\textcolor{black}{}
\begin{table}
\begin{centering}
\textcolor{black}{\caption{Dichotomous choices in experiments 5 and 6. \textcolor{black}{Each
respondent was asked to make two choices ($q_{1}\&q_{2}$ or $q_{2}\&q_{3}$
or $q_{3}\&q_{4}$ or $q_{4}\&q_{1}$), randomly and independently
assigned to this respondent in each experiment.}\label{tab:Dichotomous-choices-5-6}}
}%
\begin{tabular}{|l|c|c|c|c|}
\hline 
 & \textcolor{black}{$q_{1}$} & \textcolor{black}{$q_{2}$} & \textcolor{black}{$q_{3}$} & \textcolor{black}{$q_{4}$}\tabularnewline
\hline 
\hline 
\multirow{2}{*}{\textcolor{black}{5. Directions}} & \textcolor{black}{West\textemdash East fork} & \textcolor{black}{NorthWest\textemdash SouthEast fork} & \textcolor{black}{North\textemdash South fork} & \textcolor{black}{NorthEast\textemdash SouthWest fork}\tabularnewline
 & \textcolor{black}{$\leftarrow\ \ \text{\text{{or}}\ \ \ensuremath{\rightarrow}}$} & \textcolor{black}{$\nwarrow\ \ \text{\text{{or}}\ \ \ensuremath{\searrow}}$} & \textcolor{black}{$\uparrow\ \ \text{\text{{or}}\ \ \ensuremath{\downarrow}}$} & \textcolor{black}{$\nearrow\ \ \text{\text{{or}}\ \ \ensuremath{\swarrow}}$}\tabularnewline
\hline 
\multirow{2}{*}{\textcolor{black}{6. Colored figures}} & \textcolor{black}{one of} & \textcolor{black}{one of} & \textcolor{black}{one of} & \textcolor{black}{one of}\tabularnewline
 & \textcolor{black}{\includegraphics[scale=0.5]{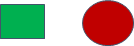}} & \textcolor{black}{\includegraphics[scale=0.5]{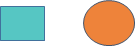}} & \textcolor{black}{\includegraphics[scale=0.5]{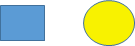}} & \textcolor{black}{\includegraphics[scale=0.5]{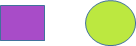}}\tabularnewline
\hline 
\end{tabular}
\par\end{centering}
\raggedright{}\textcolor{black}{For each choice $q_{i},$ the response
encoded by $+1$ is the one on the left: e.g., for $q_{1}$ in Experiment
5, the response $\leftarrow$ was encoded by $+1$.}
\end{table}
\par\end{center}

\subsubsection*{\textcolor{black}{Experiments 5 and 6}}

\textcolor{black}{In experiments 5 and 6, in each context, the character
Alice was faced with two choices out of a set of four. In all other
respects the procedure was similar to that in Experiments 1-4. The
participant was asked to select a pair of responses that respected
the character's preferences as stated in the instructions. The following
depicts the situations presented, while table \ref{tab:Dichotomous-choices-5-6}
summarizes the sets of dichotomous choices.}
\begin{description}
\item [{\textcolor{black}{Experiment\ ``Directions.''}}] \textcolor{black}{Alice
goes for a walk, and has to choose path directions at forks. Alice
wants the two directions to be as similar as possible (i.e., the angle
between them to be as small as possible). }
\item [{\textcolor{black}{Experiment\ ``Colored\ figures.''}}] \textcolor{black}{Alice
is taking a drawing lesson, and is presented with two pairs consisting
of a square and a circle (the pairs being labeled as ``Section 1''
and ``Section 2''). Alice needs to choose one figure from each section,
and she wants the two figures chosen to be of similar color. }
\end{description}

\section{\textcolor{black}{Results}}

\textcolor{black}{In Experiments 1-4, irrespective of the specific
content of the questions, there were three dichotomous choices, $q_{1},q_{2},q_{3}$,
offered to the respondents two at a time. Denoting, for each of the
choices, one of the response options $+1$ and the other $-1$, the
results have the following form:
\begin{equation}
\begin{array}{c}
\begin{array}{l|c|c|c}
c_{1} & R_{2}^{1}=1 & R_{2}^{1}=-1\\
\hline R_{1}^{1}=1 & 0 & p_{1} & p_{1}\\
\hline R_{1}^{1}=-1 & 1-p_{1} & 0 & 1-p_{1}\\
\hline  & 1-p_{1} & p_{1}
\end{array}\quad\begin{array}{l|c|c|c}
c_{2} & R_{3}^{2}=1 & R_{3}^{2}=-1\\
\hline R_{2}^{2}=1 & 0 & p_{2} & p_{2}\\
\hline R_{2}^{2}=-1 & 1-p_{2} & 0 & 1-p_{2}\\
\hline  & 1-p_{2} & p_{2}
\end{array}\\
\\
\begin{array}{l|c|c|c}
c_{3} & R_{1}^{3}=1 & R_{1}^{3}=-1\\
\hline R_{3}^{3}=1 & 0 & p_{3} & p_{3}\\
\hline R_{3}^{3}=-1 & 1-p_{3} & 0 & 1-p_{3}\\
\hline  & 1-p_{3} & p_{3}
\end{array}
\end{array}\label{eq: 3 probs}
\end{equation}
In reference to the CbD criterion (\ref{eq: criterion})-(\ref{eq: delta}),
it follows that in these experiments 
\begin{equation}
s_{odd}\left(\mathsf{E}\left[R_{1}^{1}R_{2}^{1}\right],\mathsf{E}\left[R_{2}^{2}R_{3}^{2}\right],\mathsf{E}\left[R_{3}^{3}R_{1}^{3}\right]\right)=s_{odd}\left(-1,-1,-1\right)=3,
\end{equation}
so that $D$ in (\ref{eq: criterion}) is
\begin{equation}
D=2-\Delta,
\end{equation}
where
\begin{equation}
\begin{array}{r}
\text{\ensuremath{\Delta=}}\left|\mathsf{E}\left[R_{1}^{1}\right]-\mathsf{E}\left[R_{1}^{3}\right]\right|+\left|\mathsf{E}\left[R_{2}^{2}\right]-\mathsf{E}\left[R_{2}^{1}\right]\right|+\left|\mathsf{E}\left[R_{3}^{3}\right]-\mathsf{E}\left[R_{3}^{2}\right]\right|\\
=2\left|p_{1}+p_{3}-1\right|+2\left|p_{2}+p_{1}-1\right|+2\left|p_{3}+p_{2}-1\right|.
\end{array},\label{eq: Delta in p's 3}
\end{equation}
Table \ref{tab:Probability-estimates-1-4} presents the observed values
of $\hat{p}_{1},\hat{p}_{2}$ and $\hat{p}_{3}$ for each context
of each of Experiments 1-4, and the corresponding numbers of participants
from which these probabilities were estimated.}
\begin{center}
\textcolor{black}{}
\begin{table}
\begin{centering}
\textcolor{black}{\caption{Probability estimates $\hat{p}_{1},\hat{p}_{2},\hat{p}_{3}$ that
determine the outcomes of Experiments 1-4 in accordance with (\ref{eq: 3 probs}),
and the sizes $N_{1},N_{2},N_{3}$ of the samples from which these
estimates were computed.\label{tab:Probability-estimates-1-4}}
}
\par\end{centering}
\centering{}\textcolor{black}{}%
\begin{tabular}{|l|c|c|c|c|c|c|}
\hline 
\multirow{2}{*}{\textcolor{black}{Experiment}} & \multicolumn{2}{c|}{\textcolor{black}{$c_{1}$}} & \multicolumn{2}{c|}{\textcolor{black}{$c_{2}$}} & \multicolumn{2}{c|}{\textcolor{black}{$c_{3}$}}\tabularnewline
\cline{2-7} 
 & \textcolor{black}{$\hat{p}_{1}$} & \textcolor{black}{$N_{1}$} & \textcolor{black}{$\hat{p}_{2}$} & \textcolor{black}{$N_{2}$} & \textcolor{black}{$\hat{p}_{3}$} & \textcolor{black}{$N_{3}$}\tabularnewline
\hline 
\textcolor{black}{1. Meals} & \textcolor{black}{$0.349$} & \textcolor{black}{$2090$} & \textcolor{black}{$0.658$} & \textcolor{black}{$2052$} & \textcolor{black}{$0.653$} & \textcolor{black}{$2050$}\tabularnewline
\hline 
\textcolor{black}{2. Clothes} & \textcolor{black}{$0.639$} & \textcolor{black}{$1996$} & \textcolor{black}{$0.566$} & \textcolor{black}{$2086$} & \textcolor{black}{$0.435$} & \textcolor{black}{$2110$}\tabularnewline
\hline 
\textcolor{black}{3. Presents} & \textcolor{black}{$0.547$} & \textcolor{black}{$2081$} & \textcolor{black}{$0.387$} & \textcolor{black}{$2052$} & \textcolor{black}{$0.515$} & \textcolor{black}{$2059$}\tabularnewline
\hline 
\textcolor{black}{4. Exercises} & \textcolor{black}{$0.590$} & \textcolor{black}{$2058$} & \textcolor{black}{$0.306$} & \textcolor{black}{$2024$} & \textcolor{black}{$0.580$} & \textcolor{black}{$2110$}\tabularnewline
\hline 
\end{tabular}
\end{table}
\par\end{center}

\textcolor{black}{In Experiments 5 and 6 there were four dichotomous
choices, $q_{1},q_{2},q_{3},q_{4}$, and each respondent was offered
two of them, forming one of four possible contexts. Denoting, again,
for each of the choices, one of the response options $+1$ and another
$-1$, the results have the following form:
\begin{equation}
\begin{array}{c}
\begin{array}{l|c|c|c}
c_{1} & R_{2}^{1}=1 & R_{2}^{1}=-1\\
\hline R_{1}^{1}=1 & p_{1} & 0 & p_{1}\\
\hline R_{1}^{1}=-1 & 0 & 1-p_{1} & 1-p_{1}\\
\hline  & p_{1} & 1-p_{1}
\end{array}\quad\begin{array}{l|c|c|c}
c_{2} & R_{3}^{2}=1 & R_{3}^{2}=-1\\
\hline R_{2}^{2}=1 & p_{2} & 0 & p_{2}\\
\hline R_{2}^{2}=-1 & 0 & 1-p_{2} & 1-p_{2}\\
\hline  & p_{2} & 1-p_{2}
\end{array}\\
\\
\begin{array}{l|c|c|c}
c_{3} & R_{4}^{3}=1 & R_{4}^{3}=-1\\
\hline R_{3}^{3}=1 & p_{3} & 0 & p_{3}\\
\hline R_{3}^{3}=-1 & 0 & 1-p_{3} & 1-p_{3}\\
\hline  & p_{3} & 1-p_{3}
\end{array}\quad\begin{array}{l|c|c|c}
c_{4} & R_{1}^{4}=1 & R_{1}^{4}=-1\\
\hline R_{4}^{4}=1 & 0 & p_{4} & p_{4}\\
\hline R_{4}^{4}=-1 & 1-p_{4} & 0 & 1-p_{4}\\
\hline  & 1-p_{4} & p_{4}
\end{array}
\end{array}\label{eq: 4 probs}
\end{equation}
In reference to the CbD criterion (\ref{eq: criterion})-(\ref{eq: delta}),
it follows that in these experiments 
\begin{equation}
s_{odd}\left(\mathsf{E}\left[R_{1}^{1}R_{2}^{1}\right],\mathsf{E}\left[R_{2}^{2}R_{3}^{2}\right],\mathsf{E}\left[R_{3}^{3}R_{4}^{3}\right],\mathsf{E}\left[R_{4}^{4}R_{1}^{4}\right]\right)=s_{odd}\left(1,1,1,-1\right)=4,
\end{equation}
whence, once again, 
\begin{equation}
D=2-\Delta,\label{eq: D for 4}
\end{equation}
where
\begin{equation}
\begin{array}{r}
\text{\ensuremath{\Delta=}}\left|\mathsf{E}\left[R_{1}^{1}\right]-\mathsf{E}\left[R_{1}^{4}\right]\right|+\left|\mathsf{E}\left[R_{2}^{1}\right]-\mathsf{E}\left[R_{2}^{2}\right]\right|+\left|\mathsf{E}\left[R_{3}^{2}\right]-\mathsf{E}\left[R_{3}^{3}\right]\right|+\left|\mathsf{E}\left[R_{4}^{3}\right]-\mathsf{E}\left[R_{4}^{4}\right]\right|\\
\ensuremath{=}2\left|p_{1}+p_{4}-1\right|+2\left|p_{2}-p_{1}\right|+2\left|p_{3}-p_{2}\right|+2\left|p_{4}-p_{3}\right|.
\end{array}\label{eq: Delta in p's 4}
\end{equation}
Table \ref{tab:Probability-estimates-5-6} presents the observed values
of $\hat{p}_{1},\hat{p}_{2},\hat{p}_{3},\hat{p}_{4}$ in Experiment
5 and 6, and the corresponding numbers of participants from which
these probabilities were estimated.}
\begin{center}
\textcolor{black}{}
\begin{table}
\begin{centering}
\textcolor{black}{\caption{Probability estimates $\hat{p}_{1},\hat{p}_{2},\hat{p}_{3},\hat{p}_{4}$
that determine the outcomes of Experiments 5 and 6 in accordance with
(\ref{eq: 4 probs}), and the sizes $N_{1},N_{2},N_{3},N_{4}$ of
the samples from which these estimates were computed.\label{tab:Probability-estimates-5-6}}
}
\par\end{centering}
\begin{centering}
\textcolor{black}{}%
\begin{tabular}{|l|c|c|c|c|c|c|c|c|}
\hline 
\multirow{2}{*}{\textcolor{black}{Experiment}} & \multicolumn{2}{c|}{\textcolor{black}{$c_{1}$}} & \multicolumn{2}{c|}{\textcolor{black}{$c_{2}$}} & \multicolumn{2}{c|}{\textcolor{black}{$c_{3}$}} & \multicolumn{2}{c|}{\textcolor{black}{$c_{4}$}}\tabularnewline
\cline{2-9} 
 & \textcolor{black}{$\hat{p}_{1}$} & \textcolor{black}{$N_{1}$} & \textcolor{black}{$\hat{p}_{2}$} & \textcolor{black}{$N_{2}$} & \textcolor{black}{$\hat{p}_{^{3}}$} & \textcolor{black}{$N_{3}$} & \textcolor{black}{$\hat{p}_{4}$} & \textcolor{black}{$N_{4}$}\tabularnewline
\hline 
\textcolor{black}{5. Directions} & \textcolor{black}{$0.471$} & \textcolor{black}{$1549$} & \textcolor{black}{$0.706$} & \textcolor{black}{$1504$} & \textcolor{black}{$0.645$} & \textcolor{black}{$1537$} & \textcolor{black}{$0.750$} & \textcolor{black}{$1602$}\tabularnewline
\hline 
\textcolor{black}{6. Colored figures} & \textcolor{black}{$0.419$} & \textcolor{black}{$1603$} & \textcolor{black}{$0.819$} & \textcolor{black}{$1589$} & \textcolor{black}{$0.360$} & \textcolor{black}{$1482$} & \textcolor{black}{$0.154$} & \textcolor{black}{$1517^{*}$}\tabularnewline
\hline 
\end{tabular}
\par\end{centering}
\textcolor{black}{{*} One participant assigned to context $c_{4}$
was excluded from Experiment 6 because she or he did not complete
the responses in accordance with the instructions.}
\end{table}
\par\end{center}

\textcolor{black}{Table \ref{tab:Results-of-contextuality} shows
the estimated values of $D=2-\Delta$ in all our experiments. We see
that contextuality is observed in Experiments 1-4 and 5. Experiment
6, however, shows no contextuality: the negative value in the bottom
row indicates that direct influences here are all one needs to account
for the results.}
\begin{center}
\textcolor{black}{}
\begin{table}
\begin{centering}
\textcolor{black}{\caption{\textcolor{black}{Estimated values of $D=2-\Delta$ in Experiments
1-4 ($n=3$) and 5-6 ($n=4$). Positive (negative) values of $D$
indicate contextuality (resp., noncontextuality).}\label{tab:Results-of-contextuality}}
}
\par\end{centering}
\centering{}\textcolor{black}{}%
\begin{tabular}{|r|r|r|r|r|r|r|}
\hline 
 & \textcolor{black}{1. Meals} & \textcolor{black}{2. Clothes} & \textcolor{black}{3. Presents} & \textcolor{black}{4. Exercises} & \textcolor{black}{5. Directions} & \textcolor{black}{6. Colored figures}\tabularnewline
\hline 
\hline 
\textcolor{black}{$\hat{D}=2-\hat{\Delta}$} & \textcolor{black}{$1.361$} & \textcolor{black}{$1.440$} & \textcolor{black}{$1.548$} & \textcolor{black}{$1.223$} & \textcolor{black}{$0.758$} & \textcolor{black}{$-0.984$}\tabularnewline
\hline 
\end{tabular}
\end{table}
\par\end{center}

\textcolor{black}{We evaluate statistical reliability of these results
in two ways. The first way is to compute an upper bound for the standard
deviation of $\hat{D}$ and use it to conservatively test the null-hypothesis
$D=0$ (the maximal noncontextual value) against $D>0$ (contextuality).
In Experiment 6 the alternative hypotheses changes to $D<0$ (noncontextuality),
with $D=0$ in the null hypothesis interpreted as the infimum of contextual
values. We begin by observing that each $\hat{p}_{i}$ has variance
$\frac{p\left(1-p\right)}{N_{i}}\leq\frac{1}{4N_{i}}\leq\frac{1}{4N_{min}}$,
where $N_{min}$ is the smallest among $N_{i}$ for a given experiment,
as shown in Tables \ref{tab:Probability-estimates-1-4} and \ref{tab:Probability-estimates-5-6}.
Using the independent coupling of stochastically unrelated $\hat{p_{i}}$'s,
commonly adopted in statistics, each summand in (\ref{eq: Delta in p's 3})
and (\ref{eq: Delta in p's 4}) has a variance bounded by $\frac{2}{N_{min}}$.
The different summands are not independent, but the standard deviation
of the sum cannot exceed the sum of their standard deviations. This
means that $3\sqrt{\frac{2}{N_{min}}}$ for Experiments 1-4 and $4\sqrt{\frac{2}{N_{min}}}$
for Experiments 5-6 are upper bounds for the standard deviation of
$\hat{D}$. These values are reported in Table \ref{tab: Statistical-significance}.
If we assume applicability of the central limit theorem, given the
very large sample sizes, the t-distribution-based p-values are essentially
zero. If we make no assumptions, the maximally conservative p-values
based on Chebyshev's inequality are still below the conventional significance
levels.}

\textcolor{black}{}
\begin{table}
\textcolor{black}{\caption{Statistical significance of contextuality in Experiment 1-5 and of
noncontextuality in Experiment 6. \label{tab: Statistical-significance}}
}
\centering{}\textcolor{black}{}%
\begin{tabular}{|r|r|r|r|r|r|r|}
\hline 
\textcolor{black}{Experiment:} & \textcolor{black}{1. Meals} & \textcolor{black}{2. Clothes} & \textcolor{black}{3. Presents} & \textcolor{black}{4. Exercises} & \textcolor{black}{5. Directions} & \textcolor{black}{6. Colored}\tabularnewline
\hline 
\hline 
\textcolor{black}{$\hat{D}=2-\hat{\Delta}$} & \textcolor{black}{$1.361$} & \textcolor{black}{$1.440$} & \textcolor{black}{$1.548$} & \textcolor{black}{$1.223$} & \textcolor{black}{$0.758$} & \textcolor{black}{$-0.984$}\tabularnewline
\hline 
\textcolor{black}{$N_{min}$} & \textcolor{black}{2050} & \textcolor{black}{1996} & \textcolor{black}{2052} & \textcolor{black}{2024} & \textcolor{black}{1504} & \textcolor{black}{1482}\tabularnewline
\hline 
\textcolor{black}{Upper bound for st. dev. of $\hat{D}$} & \textcolor{black}{0.094} & \textcolor{black}{0.095} & \textcolor{black}{0.094} & \textcolor{black}{0.095} & \textcolor{black}{0.146} & \textcolor{black}{0.147}\tabularnewline
\hline 
\textcolor{black}{Number of st. dev. from zero} & \textcolor{black}{$>14.5$} & \textcolor{black}{$>15.1$} & \textcolor{black}{$>16.5$} & \textcolor{black}{$>12.9$} & \textcolor{black}{$>5.1$} & \textcolor{black}{$>6.6$}\tabularnewline
\hline 
\textcolor{black}{t-distribution p-value} & \textcolor{black}{$<10^{-45}$} & \textcolor{black}{$<10^{-48}$} & \textcolor{black}{$<10^{-57}$} & \textcolor{black}{$<10^{-36}$} & \textcolor{black}{$<10^{-6}$} & \textcolor{black}{$<10^{-10}$}\tabularnewline
\hline 
\textcolor{black}{Chebyshev p-value} & \textcolor{black}{$<0.005$} & \textcolor{black}{$<0.005$} & \textcolor{black}{$<0.004$} & \textcolor{black}{$<0.006$} & \textcolor{black}{$<0.038$} & \textcolor{black}{$<0.023$}\tabularnewline
\hline 
\end{tabular}
\end{table}
\textcolor{black}{In our second statistical analysis, we computed
bootstrap distributions and constructed the $99.99\%$ bootstrap confidence
intervals for $D$ from $500000$ independent resamples for each context
of each experiment (Davison \& Hinkley, 1997). These are presented
in Figure \ref{fig:Histrograms}. As we see, the left endpoints of
the confidence intervals for experiments 1-5 are well above zero.
For experiment 6, the $99.99\%$ bootstrap confidence interval (Fig.
\ref{fig:Histrograms}) has the right endpoint well below zero, indicating
reliable lack of contextuality.}
\begin{center}
\textcolor{black}{}
\begin{figure}
\begin{centering}
\textcolor{black}{\includegraphics[scale=0.5]{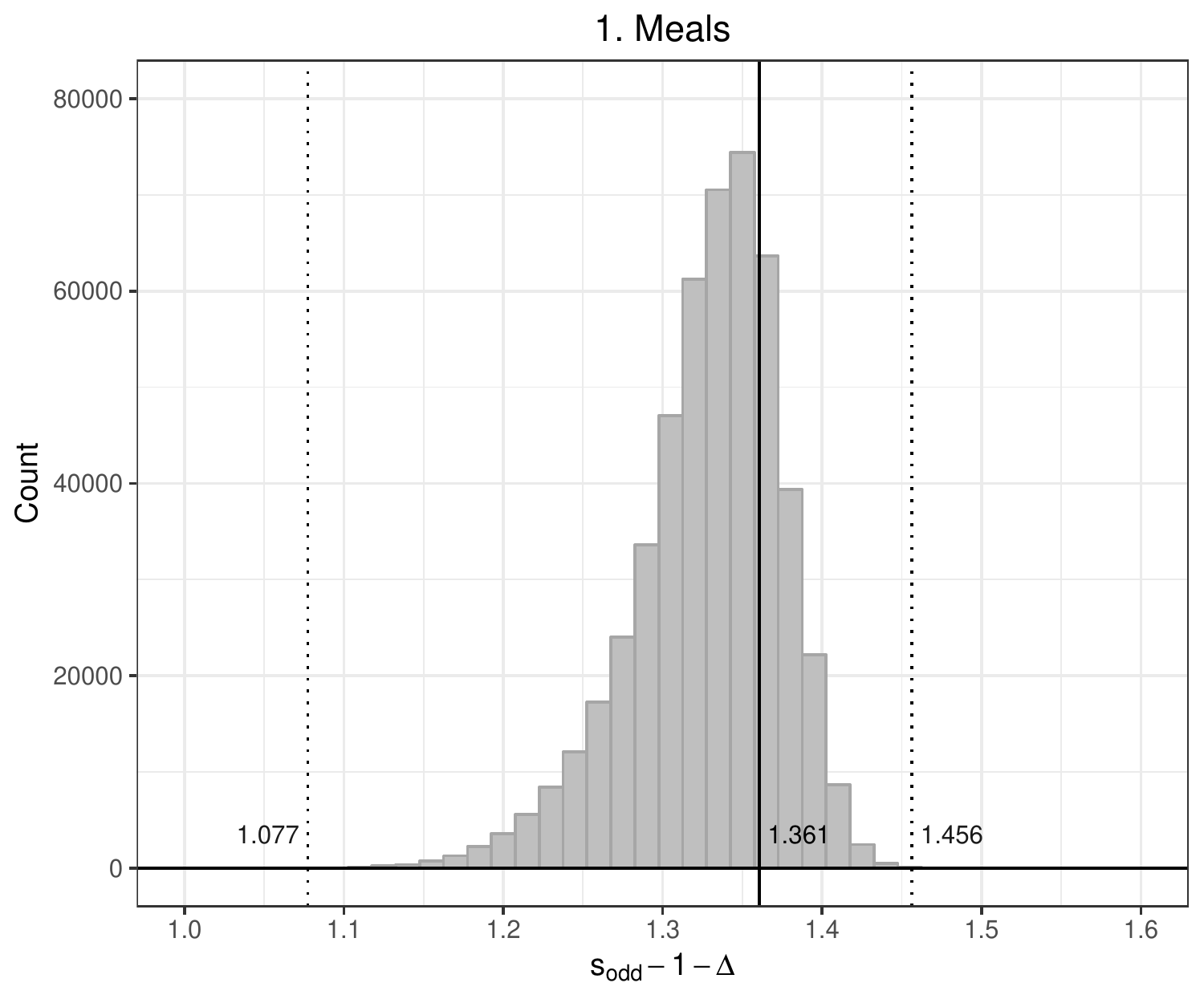} \includegraphics[scale=0.5]{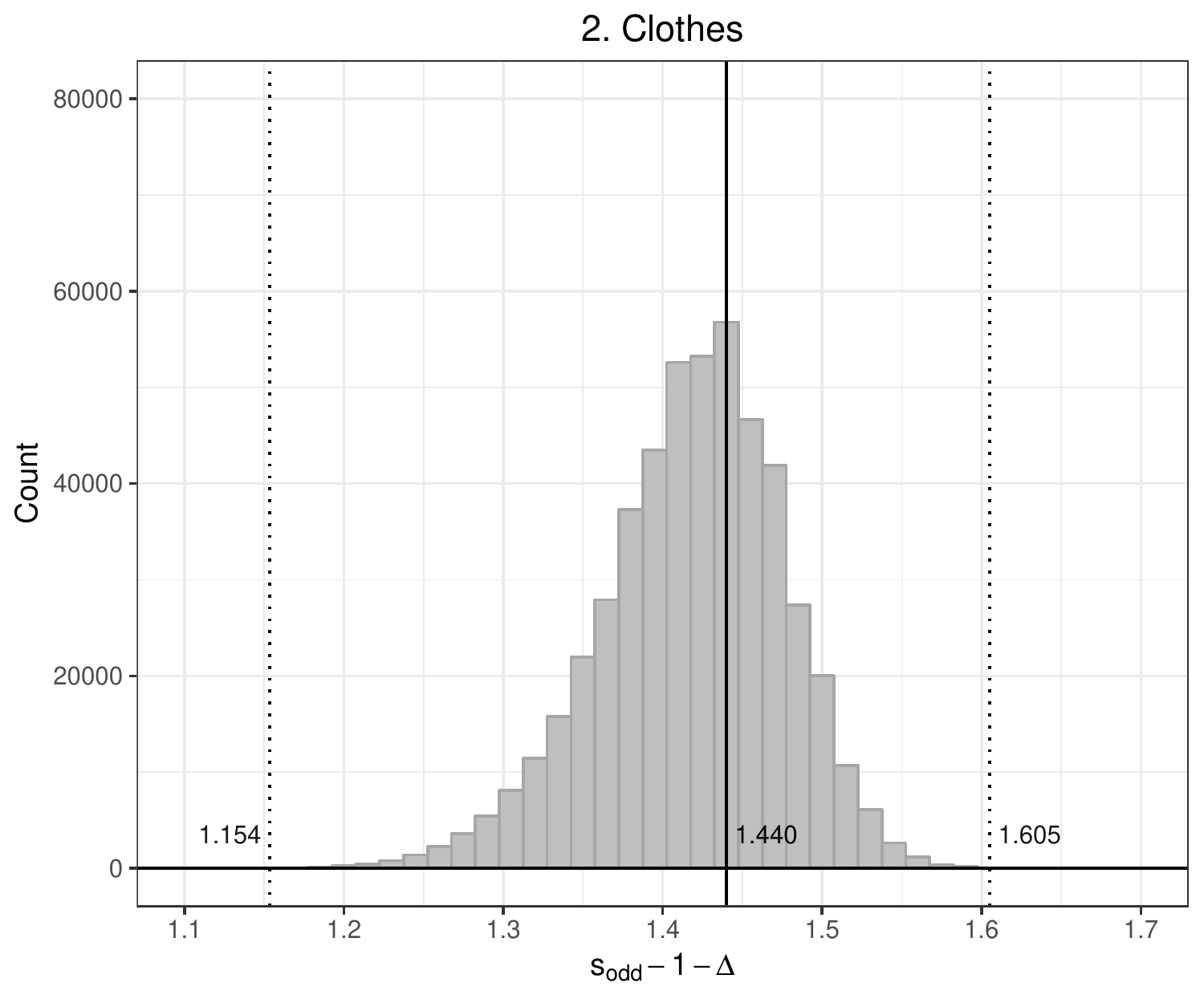}}
\par\end{centering}
\begin{centering}
\textcolor{black}{\includegraphics[scale=0.5]{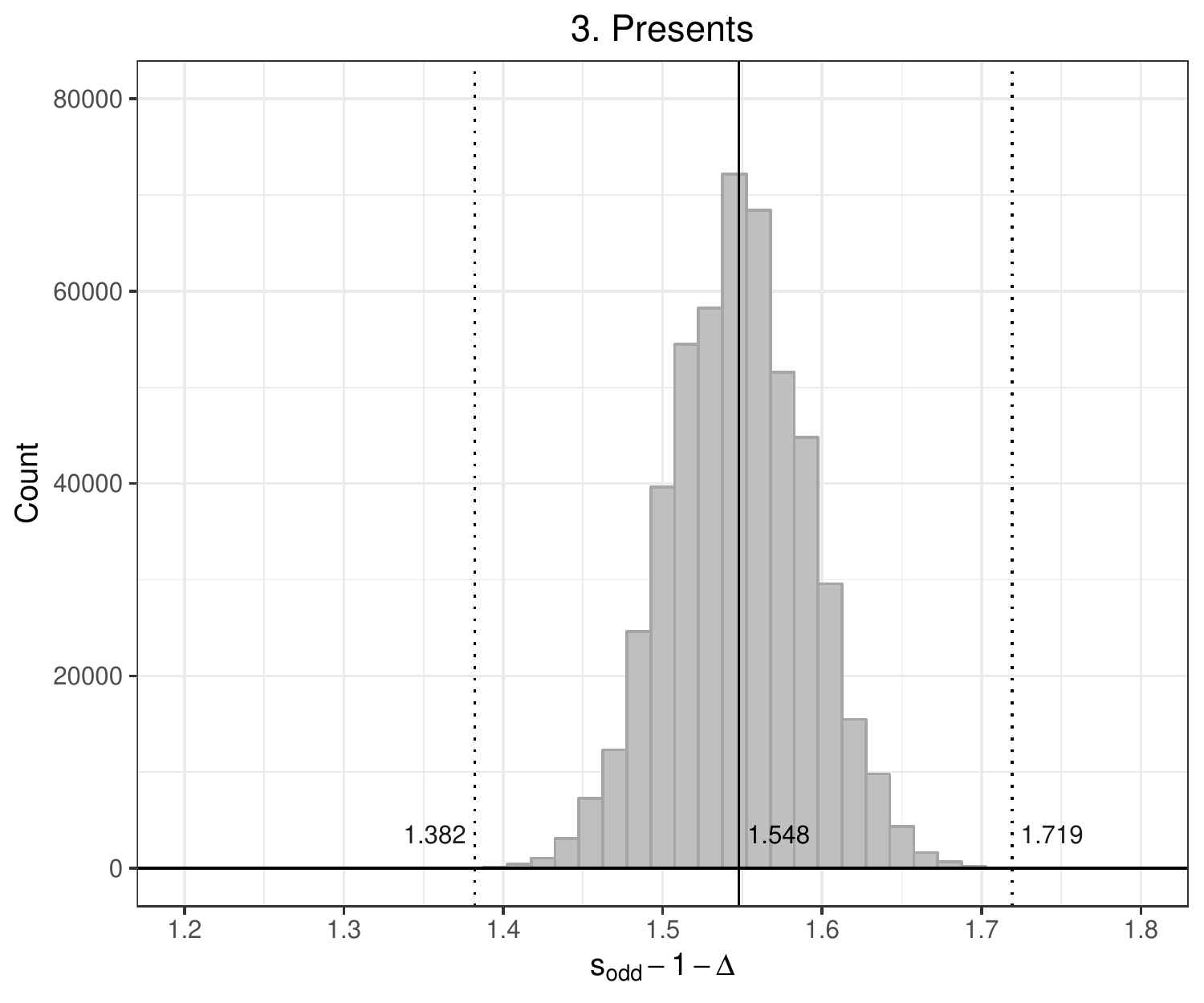} \includegraphics[scale=0.5]{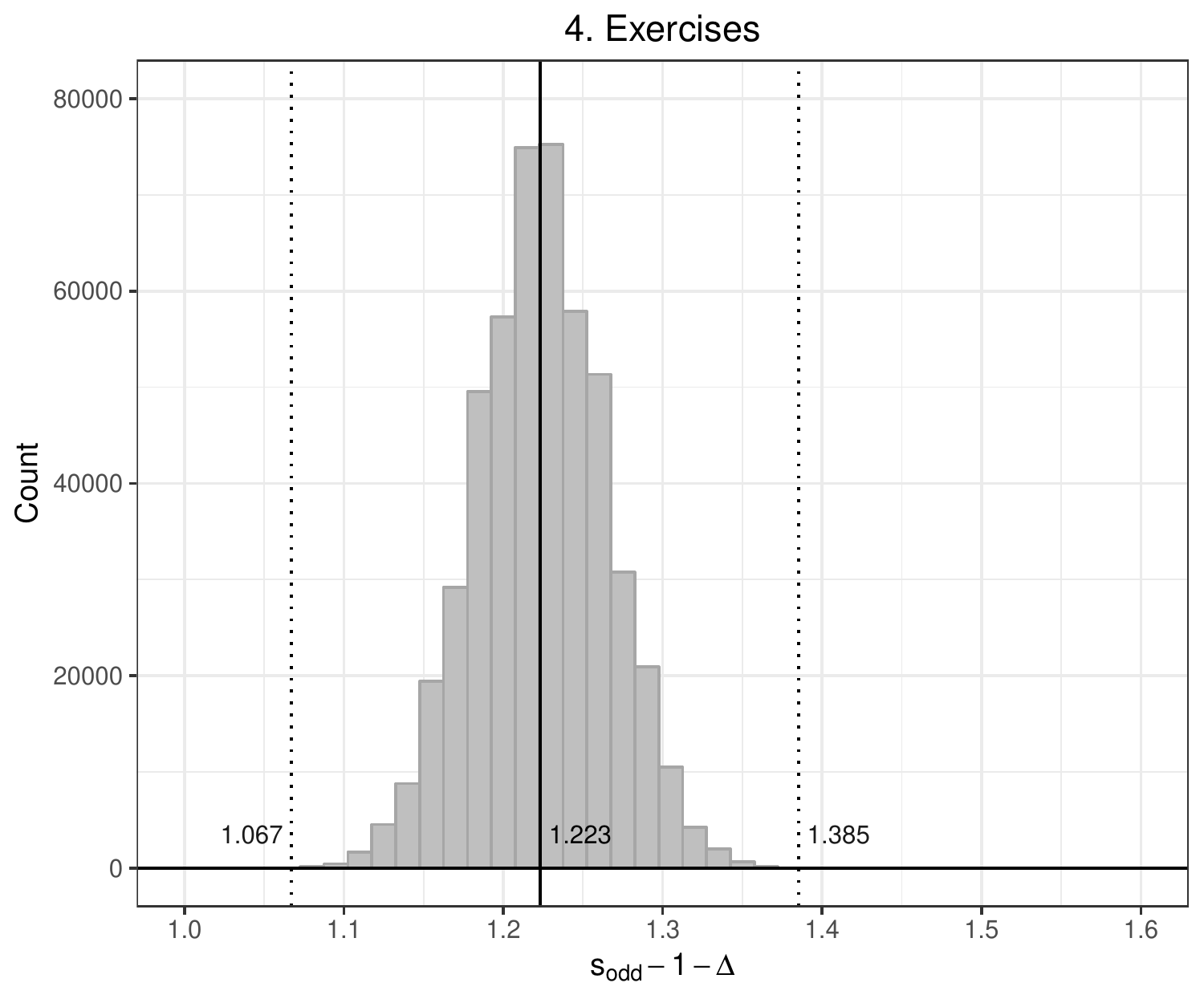}}
\par\end{centering}
\begin{centering}
\textcolor{black}{\includegraphics[scale=0.5]{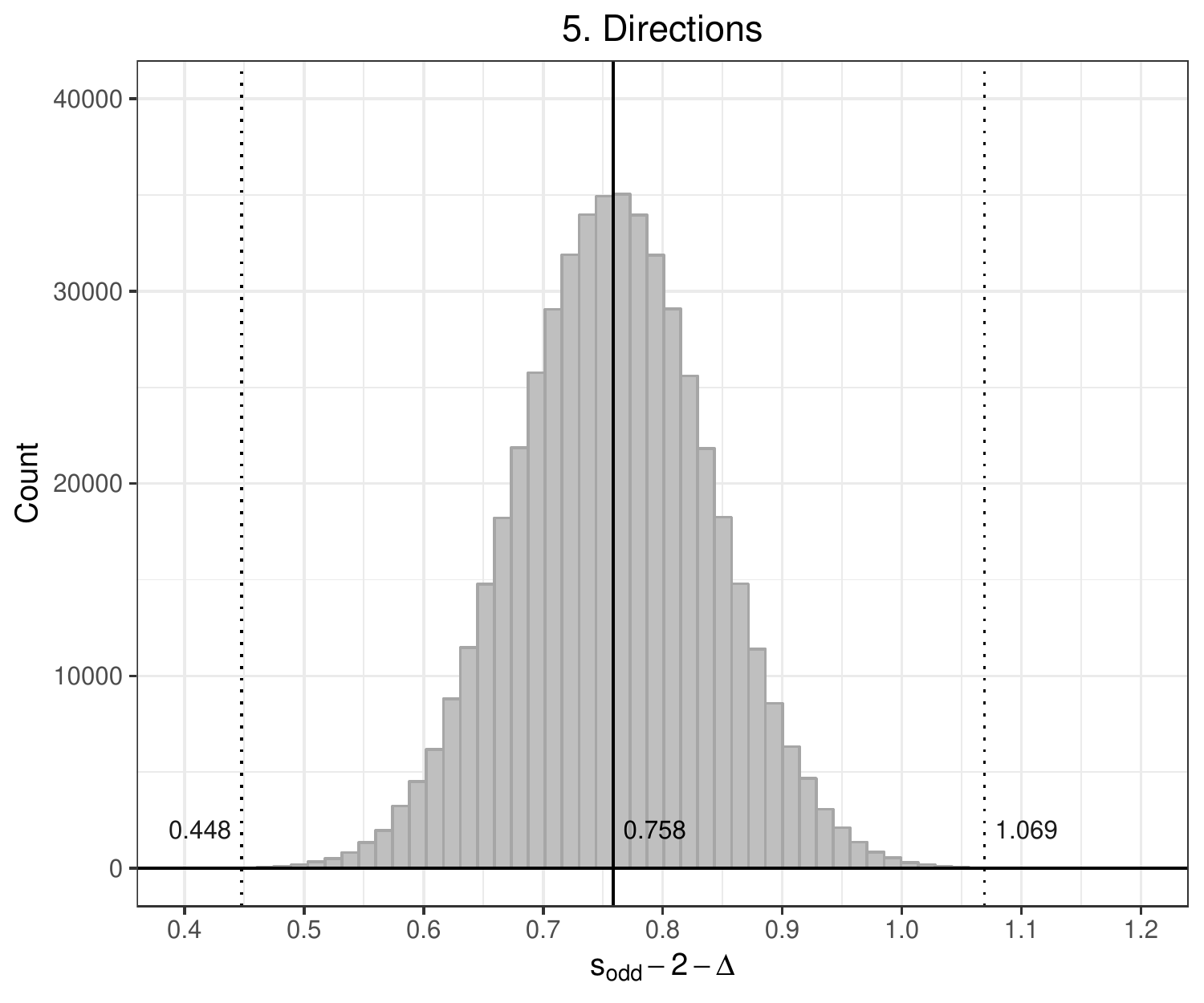} \includegraphics[scale=0.5]{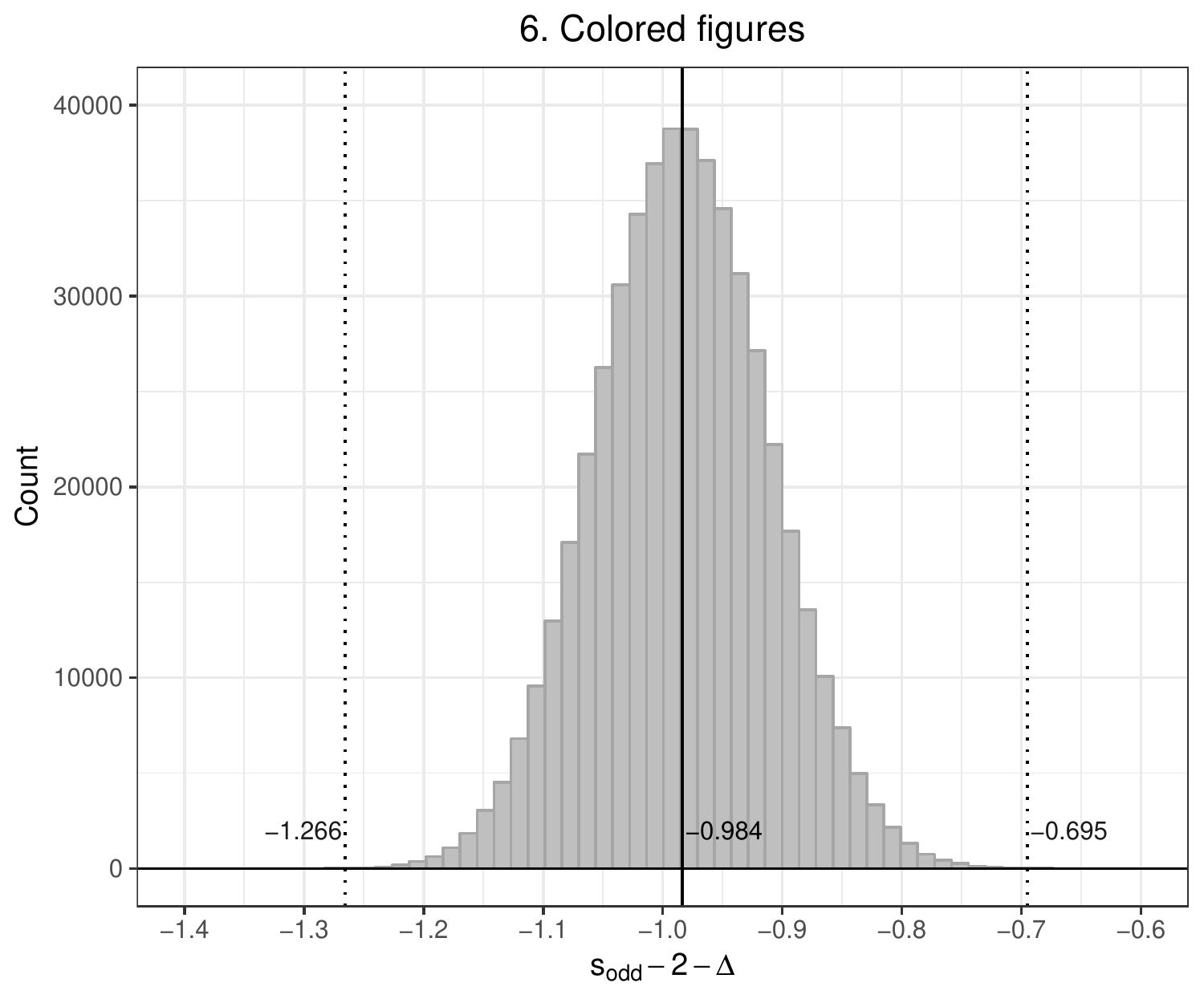}}
\par\end{centering}
\centering{}\textcolor{black}{\caption{Histograms of the bootstrap values of \textcolor{blue}{${\color{black}\hat{D}=2-\hat{\Delta}}$}
for Experiments 1-6. The solid vertical line indicates the location
of the observed sample value. The vertical dotted lines indicate the
locations of the $99.99\%$ bootstrap confidence intervals.\label{fig:Histrograms}}
}
\end{figure}
\par\end{center}

\section{\textcolor{black}{\label{sec: Discussion}Discussion}}

\textcolor{black}{Our results confirm beyond doubt the presence of
true contextuality, separated from direct influences, in simple decision
making. Compared to the ``Snow Queen'' experiment (Cervantes \&
Dzhafarov, 2018), where the paired choices belonged to different categories
(choice of characters, such as ``Gerda or Troll,'' was paired with
the choice of characteristics, such as ``kind or evil''), in our
experiments the paired choices belonged to the same category (e.g.,
two levels of arm exercises were paired with two levels of leg exercises).
The fact that our results are similar to those of the ``Snow Queen''
experiment shows that this difference is immaterial. What is material
is the design that ensures a very large value of $s_{odd}$ in the
contextuality criterion (\ref{eq: criterion}). In our experiments
it was in fact the largest possible value, one equal to the rank of
the cyclic system, $n$. This value in all but one of our experiments
was sufficient to ``beat'' direct influences, measured by $\Delta$
(in the sense that their difference exceeded $n-2$). The one exception
we got, with ``Colored figures,'' is also valuable, as it shows
that the presence of true contextuality in our experiment is an empirical
finding rather than mathematical consequence of the design: even with
$s_{odd}$ maximal in value, direct influences may very well exceed
the value of $s_{odd}-\left(n-2\right)$, making the the value of
$D$ in (\ref{eq: criterion}) negative. }

\textcolor{black}{As explained in Cervantes and Dzhafarov (2018),
in much greater detail than in the present brief recap, it is important
that the design we used was between-subjects, i.e. each respondent
in each experiment was assigned to a single context only. The reason
for this is that if a single respondent were asked to make pairs of
choices in all three contexts (in Experiments 1-4) or in all four
contexts (in Experiments 5 and 6), it would have created an empirical
joint distribution of all the random variables in the respective systems.
This would contravene the logic of CbD, in which different contexts
are mutually exclusive, and the random variables in different rows
of content-context matrices are stochastically unrelated (have no
joint distribution).}

\textcolor{black}{One might question another aspect of our experimental
design: the fact that the respondents were not allowed to contravene
their instructions and make incorrect choices (e.g., choose two ``high''
options or two ``low'' options in Experiments 1-4). The main reason
for this is that in a crowdsourcing experiment, with no additional
information about the respondents, it is difficult to understand what
could lead a person not to follow the simple instructions. Ideally,
one would want to separate data due to deliberate non-compliance or
disregard from ``honest mistakes,'' and this is impossible. In fact,
it is hard to fathom what an ``honest mistake'' in a situation as
simple as ours might be. In the ``Snow Queen'' experiment (Cervantes
and Dzhafarov, 2018), where the choices were, arguably, less simple
than in the present experiments, incorrect responses were allowed,
and their percentage was just over 8\%. Their inclusion or exclusion
did not make any difference for analysis and conclusions. }

\textcolor{black}{In the opening of the paper and at the end of Section
\ref{subsec: A-numerical-example} we alluded to the interpretation
of true contextuality in terms of the ``wholes'' irreducible to
interacting ``parts.'' One must not mistake this interpretation
for the old adage that ``the whole is something besides the parts''
(Aristotle) or, as reformulated by Kurt Koffka (1935), \textquotedblleft the
whole is something else than the sum of its parts\textquotedblright{}
(p.176). These and similar statements are not only vague, they have
also been rendered essentially meaningless by their indiscriminate
application to all kinds of situations. In most of cases one has a
justifiable suspicion that what is meant is that parts interact, or
that someone can discern a pattern in them. This is probably always
true when the parts are deterministic entities. In the case of random
variables, however, there is a rigorous analytic meaning of saying
that the whole is different from, and indeed greater than a system
of parts with all their interactions. Random variables measuring or
responding to one and the same ``part'' (property or stimulus) have
different identities in different ``wholes'' (contexts), with the
difference being greater than warranted by the mere distributional
differences caused by their interactions with other elements of the
``wholes.'' If this sounds too philosophical to be of importance
in scientific practice, we have an example of quantum mechanics to
counter this view. }

\textcolor{black}{Contextuality in quantum mechanics is not a predictive
theory, and it is never used to derive any parts of quantum-mechanical
theory. Rather the other way around, quantum-mechanical theory is
used to determine a system's behavior, from which it is possible to
establish if the system is contextual. Thus, in the most famous example
of quantum contextuality, involving spins of entangled particles (Bell,
1964, 1966), the correlations between spins are computed by standard
quantum-theoretic formulas, and the results are used to establish
that, for certain choices of axes along which the spins are measured,
the system is contextual. The computations themselves make no use
of contextuality, nor are they being amended in any way as a result
of establishing contextuality or lack thereof. Nevertheless, the contextuality
analysis of spins of entangled particles (Bell, 1964, 1966; Clauser,
Horne, Shimony, \& Holt, 1969; Fine 1982), mathematically related
to a special case of our contextuality criterion (\ref{eq: criterion}),
with $n=4$ and $\Delta=0$, is considered highly significant. A prominent
experimental physicist, Alain Aspect, called it ``one of the profound
discoveries of the {[}20th{]} century'' (Aspect, 1999), and teams
of experimentalists have put much effort into verifying that the quantum-mechanical
predictions used to derive it are correct (Handsteiner et al., 2017).
The reason for this is, of course, that contextuality reveals something
about one of the most fundamental aspects of quantum theory: the nature
of random variables used to describe quantum phenomena. Thus, it is
significant that typical systems of random variables describing classical
mechanics happen to be noncontextual, while some quantum-mechanical
systems are contextual. In time it has also become clear that, in
addition to its foundational significance, quantum contextuality correlates
with physical properties that can be used for practical purposes.
Physicists and computer scientists at present are beginning to pose
the question of ``contextuality advantage'' or ``contextuality
as a resource,'' which is the question of whether contextuality or
noncontextuality of a system can be utilized for practical purposes.
It is argued, e.g., that the degree of contextuality (a notion we
have not discussed in this paper, see Dzhafarov, Cervantes \& Kujala,
2017; Kujala \& Dzhafarov, 2016) is directly related to computational
advantage of quantum computing over conventional one (Abramsky, Barbosa,
\& Mansfield, 2017; Frembs, Roberts, \& Bartlett, 2018).}

\textcolor{black}{Psychology shares the mandatory use of random variables
with quantum physics: stochasticity of responses in most areas of
psychology is inherent, it cannot be reduced by progressively greater
control of stimuli and conditions. The status and role of contextuality
therefore can be expected to be similar. The same as in quantum physics,
contextuality analysis is not a predictive model competing with other
models. Thus, in constructing a model to fit our data, contextuality
analysis can help only in the trivial sense: as with any other property
of the data, if contextuality or noncontextuality of them is established,
a model is to be rejected if it fails to predict this property. As
an example, one could attempt to fit our data by a model with responses
being chosen from some ``covertly'' evoked initial responses actualized
with the aid of some conflict resolution scheme. Assume that each
question $q$ has a probability $h$ of being ``covertly'' answered
$+1$ (standing here for one of the two options), and that in a context
$c=\left(q,q'\right)$ these covert responses occur independently,
so that $\left(+1,+1\right)$ occurs with probability $hh'$, $\left(+1,-1\right)$
with probability $h\left(1-h'\right)$ etc. If the combination of
covert responses is allowed by the instructions (e.g., West and North-West
in Experiment 5, or Red and Orange in Experiment 6), they turn into
observed responses; if the combination is prohibited (say, West and
South-East, or Red and Blue), the respondent randomly flips one of
the two responses, say, with probability 1/2. Then the observed probability
of choosing an allowed combination $\left(+1,-1\right)$ is computed
as $h\left(1-h'\right)+hh'/2+\left(1-h\right)\left(1-h'\right)/2$.
This model can be shown to predict that a system in our experiments
is contextual, but it is incompatible with the noncontextuality in
Experiment 6. This was only one example, however. Simple models that
can predict both contextual and noncontextual outcomes in our experiments
can be readily constructed, because all one has to predict are three
probabilities $\left(p_{1},p_{2},p_{3}\right)$ in (\ref{eq: 3 probs})
for Experiments 1-4, and four probabilities $\left(p_{1},p_{2},p_{3},p_{4}\right)$
in (\ref{eq: 4 probs}) for Experiments 5-6. Consider, e.g., a model
with eight triples $\left(+1,+1,+1\right),\left(+1,+1,-1\right),\ldots,\left(-1,-1,-1\right)$,
mental states evoked with certain probabilities, with the following
decision rule: if the context is $\left(q_{i},q_{j}\right)$, $i,j=1,2,3$
and the mental state contains $r_{i}$ ($+1$ or $-1$) and $r_{j}$
($+1$ or $-1$) in the $i$th and $j$th positions, respectively,
then respond $\left(r_{i},r_{j}\right)$ if this response combination
is allowable; if the combination is forbidden, choose one of the allowable
combinations with probability 1/2. The model has 7 free parameters,
and it can fit $\left(p_{1},p_{2},p_{3}\right)$ in Experiments 1-4
precisely. For Experiments 5 and 6, the eight triples have to be replaced
with 16 quadruples. We need not get into discussing such models here:
it was not a purpose of our experiments to achieve a deeper understanding
of how someone chooses to eat soup and beans over burger and salad.
Rather our aim was to capitalize on the psychological transparency
and modeling simplicity of such choices to firmly establish that ``quantum-like''
contextuality can be observed outside quantum physics, in human behavior.
Recall that many previous attempts to demonstrate behavioral contextuality
have failed, so our paper is only one of the first two steps (the
other one being the ``Snow Queen'' experiment in Cervantes \& Dzhafarov,
2018) on the path of identifying contextual systems in human behavior. }

\textcolor{black}{Thinking by analogy with the ``contextuality advantage''
mentioned above, can we, at this early stage of exploration, point
out any properties of human behavior as correlating with or being
indicated by contextuality? One obvious fact is that in our experiments
contextuality is negatively related to the value of $\Delta$, the
amount of direct influences. Lack of direct influences means that
the probability of choosing a particular option, say, burger, is the
same irrespective of what context this option is included in (e.g.,
whether the plain skirt is chosen in the skirt-blouse combination
or in the jacket-skirt one). The lack of direct influences would result
in the maximal possible value of $D=2$. This simplicity, however,
is specific to our design, in which $s_{odd}$ function does not vary.
For a more general class of systems of random variables, one cannot
simply replace contextuality with a measure inversely related to the
amount of direct influences (we even have examples when the two are
synergistic rather than antagonistic). Another dimension of human
behavior that can be related to contextuality can be called the degree
of ``similarity'' or ``unanimity'' of decisions across pools of
respondents, or across repeated responses by the same person when
a within-subject design is possible (as in Cervantes \& Dzhafarov,
2017a, b, and Zhang \& Dzhafarov, 2017). Consider, e.g., one of our
Experiments 1-4, and assume that the respondents agreed among themselves
on what option to choose in each context. The system then would become
deterministic and noncontextual, with $D=-4$ or $D=0$, depending
on the pattern of choices agreed upon. Small deviations from an agreed-on
pattern would result in small deviations from the corresponding values
of $D$. On the other extreme we have maximal diversity, when in each
context the opposite options are chosen with equal probabilities.
In this case the system would reach the maximal possible degree of
contextuality. Again, it is not possible to simply replace contextuality
with some measure of unanimity, such as variance: the maximal value
of contextuality can also be achieved without maximal diversity of
responses, and ``deep noncontextuality,'' with $D$ between $-4$
and $0$, can be achieved with non-deterministic systems. With due
caution, one can conjecture that the degree of (non)contextuality,
for a given format of the content-context matrix, may reflect a combination
of the two dimensions mentioned: (in)consistency of choices across
contexts (reflecting the amount of direct influences) and unanimity/diversity
of choices made in each context across a pool of respondents or repeated
in a within-subject design (reflecting the amount of determinism/stochasticity).
We will not know if this or other relations of contextuality to various
aspects of behavior can be established until we broaden our knowledge
of the degree of (non)contextuality to a much larger class of behavioral
systems.}

\section*{\textcolor{black}{REFERENCES}}

\textcolor{black}{\setlength{\parindent}{0cm}\everypar={\hangindent=15pt}}
\begin{enumerate}
\item \textcolor{black}{Abramsky, S., \& Brandenburger, A. (2011). The sheaf-theoretic
structure of non-locality and contextuality. New Journal of Physics,
13(11), 113036. \url{https://doi.org/10.1088/1367-2630/13/11/113036} }
\item \textcolor{black}{Abramsky, S., Barbosa, R., Mansfield, S. (2017).
Contextual fraction as a measure of contextuality. Physical Review
Letters 119, 050504. \url{https://doi.org/10.1103/PhysRevLett.119.050504}}
\item \textcolor{black}{Adenier, G., \& Khrennikov, A. Y. (2017). Test of
the no-signaling principle in the Hensen \textquotedblleft loophole-free
CHSH experiment.\textquotedblright{} Fortschritte der Physik, 65,
1600096. \url{https://doi.org/10.1002/prop.201600096}}
\item \textcolor{black}{Aerts, D. (2014). Quantum theory and human perception
of the macro-world. Frontiers in Psychology, 5, 1\textendash 19. \url{https://doi.org/10.3389/fpsyg.2014.00554} }
\item \textcolor{black}{Aerts, D., Gabora, L., \& Sozzo, S. (2013). Concepts
and their dynamics: A quantum-theoretic modeling of human thought.
Topics in Cognitive Science, 5(4), 737\textendash 772. \url{https://doi.org/10.1111/tops.12042}}
\item \textcolor{black}{Araújo, M., Quintino, M.T. , Budroni, C. , Cunha,
M. T., Cabello, A. (2013). All noncontextuality inequalities for the
n-cycle scenario, Physical Review A 88, 022118. \url{https://doi.org/10.1103/PhysRevA.88.022118}}
\item \textcolor{black}{Asano, M., Hashimoto, T., Khrennikov, A. Y., Ohya,
M., \& Tanaka, Y. (2014). Violation of contextual generalization of
the Leggett-Garg inequality for recognition of ambiguous figures.
Physica Scripta, T163, 14006. \url{https://doi.org/10.1088/0031-8949/2014/T163/014006} }
\item \textcolor{black}{Aspect, A. (1999). Bell's inequality tests: More
ideal than ever. Nature 398, 189-190. \url{https://doi.org/10.1038/18296}}
\item \textcolor{black}{Bell, J. S. (1964). On the Einstein-Podolsky-Rosen
paradox. Physics, 1(3), 195\textendash 200. }
\item \textcolor{black}{Bell, J. S. (1966). On the problem of hidden variables
in quantum mechanics. Review of Modern Physic 38, 447-453. \url{https://doi.org/10.1103/RevModPhys.38.447}}
\item \textcolor{black}{Bruza, P. D., Kitto, K., Nelson, D., \& McEvoy,
C. (2009). Is there something quantum-like about the human mental
lexicon? Journal of Mathematical Psychology, 53(5), 362\textendash 377.
\url{https://doi.org/10.1016/j.jmp.2009.04.004} }
\item \textcolor{black}{Bruza, P. D., Kitto, K., Ramm, B. J., \& Sitbon,
L. (2015). A probabilistic framework for analysing the compositionality
of conceptual combinations. Journal of Mathematical Psychology, 67,
26\textendash 38. \url{https://doi.org/10.1016/j.jmp.2015.06.002} }
\item \textcolor{black}{Bruza, P. D., Wang, Z., \& Busemeyer, J. R. (2015).
Quantum cognition: a new theoretical approach to psychology. Trends
in Cognitive Sciences, 19(7), 383\textendash 393. \url{https://doi.org/10.1016/j.tics.2015.05.001} }
\item \textcolor{black}{Busemeyer, J.R., \& Bruza, P.D. (2012). Quantum
Models of Cognition and Decision. Cambridge: Cambridge University
Press. \url{https://doi.org/10.1017/CBO9780511997716}}
\item \textcolor{black}{Cervantes, V. H., \& Dzhafarov, E. N. (2017a). Exploration
of contextuality in a psychophysical double-detection experiment.
In J. A. de Barros, B. Coecke, E. Pothos (Eds.), Quantum Interaction.
LNCS (Vol. 10106, pp. 182-193). Dordrecht: Springer. . \url{https://doi.org/10.1007/978-3-319-52289-0_15}}
\item \textcolor{black}{Cervantes, V. H., \& Dzhafarov, E. N. (2017b). Advanced
analysis of quantum contextuality in a psychophysical double-detection
experiment. Journal of Mathematical Psychology 79, 77-84. \url{https://doi.org/10.1016/j.jmp.2017.03.003}}
\item \textcolor{black}{Cervantes, V. H., \& Dzhafarov, E. N. (2018). Snow
Queen is evil and beautiful: Experimental evidence for probabilistic
contextuality in human choices. Decision 5, 193-204. \url{https://doi.org/10.1037/dec0000095}}
\item \textcolor{black}{Clauser, J. F., Horne, M. A., Shimony, A., \& Holt,
R. A. (1969). Proposed experiment to test local hidden-variable theories.
Physical Review Letters, 23, 880\textendash 884. \url{https://doi.org/10.1103/PhysRevLett.23.880} }
\item \textcolor{black}{CrowdFlower (2018). \url{http://www.crowdflower.com/survey}}
\item \textcolor{black}{Davison, A. C., \& Hinkley, D. V. (1997). Bootstrap
Methods and their Application (1st ed.). New York, NY, USA: Cambridge
University Press. \url{https://doi.org/10.1017/CBO9780511802843}}
\item \textcolor{black}{Dzhafarov, E.N. (2003). Selective influence through
conditional independence. Psychometrika, 68, 7-26. \url{https://doi.org/10.1007/BF02296650}}
\item \textcolor{black}{Dzhafarov, E. N., Cervantes, V. H., \& Kujala, J.
V. (2017). Contextuality in canonical systems of random variables.
Philosophical Transactions of the Royal Society A, 375, 20160389.
\url{https://doi.org/10.1098/rsta.2016.0389} }
\item \textcolor{black}{Dzhafarov, E. N., \& Kujala, J. V. (2014a). Contextuality
is about identity of random variables. Physica Scripta, 2014(T163),
14009. \url{https://doi.org/10.1088/0031-8949/2014/T163/014009} }
\item \textcolor{black}{Dzhafarov, E. N., \& Kujala, J. V. (2014b). On selective
influences, marginal selectivity, and Bell/CHSH inequalities. Topics
in Cognitive Science, 6(1), 121\textendash 128. \url{https://doi.org/10.1111/tops.12060} }
\item \textcolor{black}{Dzhafarov, E. N., \& Kujala, J. V. (2016). Context-content
systems of random variables: The Contextuality-by-Default theory.
Journal of Mathematical Psychology, 74, 11\textendash 33. \url{https://doi.org/10.1016/j.jmp.2016.04.010} }
\item \textcolor{black}{Dzhafarov, E. N., \& Kujala, J. V. (2017a). Contextuality-by-Default
2.0: Systems with Binary Random Variables. In J. A. de Barros, B.
Coecke, \& E. Pothos (Eds.), Quantum Interaction. LNCS (Vol. 10106,
pp. 16\textendash 32). Dordrecht: Springer. \url{https://doi.org/10.1007/978-3-319-52289-0_2}}
\item \textcolor{black}{Dzhafarov, E. N., \& Kujala, J. V. (2017b). Probabilistic
foundations of contextuality. Fortschritte Der Physik, 65, 1600040
\url{https://doi.org/10.1002/prop.201600040} }
\item \textcolor{black}{Dzhafarov, E. N., Kujala, J. V., Cervantes, V. H.,
Zhang, R., \& Jones, M. (2016). On contextuality in behavioural data.
Philosophical Transactions of the Royal Society A, 374, 20150234.
\url{https://doi.org/10.1098/rsta.2015.0234} }
\item \textcolor{black}{Dzhafarov, E. N., Kujala, J. V., \& Larsson, J.-Å.
(2015). Contextuality in three types of quantum-mechanical systems.
Foundations of Physics, 45(7), 762\textendash 782. \url{https://doi.org/10.1007/s10701-015-9882-9} }
\item \textcolor{black}{Dzhafarov, E. N., Zhang, R., \& Kujala, J. V. (2015).
Is there contextuality in behavioural and social systems? Philosophical
Transactions of the Royal Society A, 374, 20150099. \url{https://doi.org/10.1098/rsta.2015.0099} }
\item \textcolor{black}{Fine, A. (1982). Hidden variables, joint probability,
and the Bell inequalities. Physical Review Letters, 48(5), 291\textendash 295.
\url{https://doi.org/10.1103/PhysRevLett.48.291} }
\item \textcolor{black}{Frembs, M., Roberts, S., \& Bartlett, S. D. (2018).
Contextuality as a resource for measurement-based quantum computation
beyond qubits. \url{https://arxiv.org/abs/1804.07364}}
\item \textcolor{black}{Handsteiner, J., Friedman, A. S., Rauch, D., Gallicchio,
J., Liu, B., Hosp, H., Kofler, J., Bricher, D., Fink, M., Leung, C.,
Mark, A., Nguyen, H. T., Sanders, I., Steinlechner, F., Ursin, R.,
Wengerowsky, S., Guth, A. H., Kaiser, D. I., Scheidl, T., \& Zeilinger,
A. (2017). Cosmic Bell Test: Measurement Settings from Milky Way Stars.
Physical Review Letters, 118, 060401. \url{https://doi.org/10.1103/PhysRevLett.118.060401}}
\item \textcolor{black}{Khrennikov, A. (2009). Contextual approach to quantum
formalism. Springer, Berlin-Heidelberg-New York. \url{https://doi.org/10.1007/978-1-4020-9593-1}}
\item \textcolor{black}{Khrennikov, A. (2010). Ubiquitous quantum structure:
from psychology to finances. Springer, Berlin-Heidelberg-New York.
\url{https:doi.org/10.1007/978-3-642-05101-2}}
\item \textcolor{black}{Kochen, S., \& Specker, E. P. (1967). The problem
of hidden variables in quantum mechanics. Journal of Mathematics and
Mechanics, 17(1), 59\textendash 87. \url{http://doi.org/10.1512/iumj.1968.17.17004}}
\item \textcolor{black}{Koffka, K. (1935). Principles of Gestalt psychology.
Routledge, London, UK.}
\item \textcolor{black}{Kujala, J. V., \& Dzhafarov, E. N. (2016). Proof
of a conjecture on contextuality in cyclic systems with binary variables.
Foundations of Physics, 46, 282\textendash 299. \url{https://doi.org/10.1007/s10701-015-9964-8} }
\item \textcolor{black}{Kujala, J. V., Dzhafarov, E. N., \& Larsson, J-Å.
(2015). Necessary and sufficient conditions for an extended noncontextuality
in a broad class of quantum mechanical systems. Physical Review Letters,
115(15), 150401. \url{https://doi.org/10.1103/PhysRevLett.115.150401} }
\item \textcolor{black}{Kurzynski, P., Ramanathan, R., \& Kaszlikowski,
D. (2012). Entropic test of quantum contextuality. Physical Review
Letters, 109(2), 020404. \url{https://doi.org/10.1103/PhysRevLett.109.020404}}
\item \textcolor{black}{Liang, Y.-C., Spekkens, R. W., Wiseman, H. M. (2011).
Specker\textquoteright s parable of the overprotective seer: A road
to contextuality, nonlocality and complementarity. Physics Reports
506, 1-39. \url{https://doi.org/10.1016/j.physrep.2011.05.001}}
\item \textcolor{black}{Popescu, S., \& Rohrlich, D. (1994). Quantum nonlocality
as an axiom. Foundations of Physics, 24(3), 379\textendash 385. \url{https://doi.org/10.1007/BF02058098} }
\item \textcolor{black}{Townsend, J.T., \& Schweickert, R. (1989). Toward
the trichotomy method of reaction times: Laying the foundation of
stochastic mental networks. Journal of Mathematical Psychology, 33,
309-327. \url{https://doi.org/10.1016/0022-2496(89)90012-6}}
\item \textcolor{black}{Wang, Z., \& Busemeyer, J. R. (2013). A quantum
question order model supported by empirical tests of an a priori and
precise prediction. Topics in Cognitive Science, 5(4), 689\textendash 710.
\url{https://doi.org/10.1111/tops.12040} }
\item \textcolor{black}{Zhang, R., \& Dzhafarov, E. N. (2017). Testing contextuality
in cyclic psychophysical systems of high ranks. In J. A. de Barros,
B. Coecke, E. Pothos (Eds.) Quantum Interaction. LNCS (Vol. 10106,
pp. 151\textendash 162). Dordrecht: Springer. \url{https://doi.org/10.1007/978-3-319-52289-0_12}}
\end{enumerate}

\end{document}